# Clustering constrained on linear networks

Asael Fabian Martínez

Departamento de Matemáticas, Universidad Autónoma Metropolitana, Unidad Iztapalapa, Ciudad de México, Mexico.

`fabian@xanum.uam.mx`

Somnath Chaudhuri

Research Group on Statistics, Econometrics and Health (GRECS), University of Girona, Spain, and CIBER of Epidemiology and Public Health (CIBERESP), Spain.

`chaudhuri.somnath@udg.edu`

Carlos Díaz-Avalos

IIMAS, Universidad Nacional Autónoma de México, Ciudad Universitaria, Mexico.

`carlos@sigma.iimas.unam.mx`

Pablo Juan

IMAC, University Jaume I, Castellón, Spain.

`juan@uji.es`

Jorge Mateu

Department of Mathematics, University of Jaume I, Castellón, Spain.

`mateu@uji.es`

Ramsés H. Mena

IIMAS, Universidad Nacional Autónoma de México, Ciudad Universitaria, Mexico.

`ramses@sigma.iimas.unam.mx`

**Abstract**

An unsupervised classification method for point events occurring on a network of lines is proposed. The idea relies on the distributional flexibility and practicality of random partition models to discover the clustering structure featuring observations from a particular phenomenon taking place on a given set of edges. By incorporating the spatial effect in the random partition distribution, induced by a Dirichlet process, one is able to control the distance between edges and events, thus leading to an appealing clustering method. A Gibbs sampler algorithm is proposed and evaluated with a sensitivity analysis. The proposal is motivated and illustrated by the analysis of crime and violence patterns in Mexico City.

**Keywords:**  Bayesian nonparametrics; Penalty function; Random partition model; Spatial clustering.

# 1   Introduction

Violence and insecurity are major concerns in most Latin American countries. The reasons and causes of increasing criminality are many, therefore methodologies to study and to diminish the associated incidence rates are constantly sought. Cities like São Paulo, Managua, San Salvador and Mexico City have suffered a notorious increase in crime levels, linked both to economic factors and to corruption in the police forces Pansters & Castillo Berthier (2007). One of the most urgent demands to politicians and the mayor of Mexico City is to implement better methods for surveillance and crime control. Some first steps to deal with this necessity is the identification of areas with high crime incidence. Such hot-spots need to be identified and characterized in order to get better information for security forces, and for managers and designers of social programs to control and mitigate the causes of the high criminality in those locations.

The point pattern nature of crime occurrences suggests that a sensible approach for the analysis and modeling of crime data is based on point process theory. However, given that most of the crimes in a city are georeferenced along streets, the usual point process theory, where the events occur in a Euclidean space, is not always suitable. This is mainly due to the fact that nearest neighbor distances need to be defined along a linear network of streets, e.g. through the Manhattan distance.

Point processes in linear networks have been intensively investigated in the last decade. Nowadays, the wide spread of smartphones has increased the availability of point referenced data, which has caused a boom of research papers using point pattern data. For example, the $K$-function on a linear network has been investigated by Okabe & Yamada (2001), Yamada & Thill (2004), Ang et al. (2012), and Baddeley et al. (2017) among others. Specific modeling features, such as separability, have been investigated for example by Mateu et al. (2020) and by Gilardi et al. (2021).

A common problem that has received particular attention is how to detect clusters in the network using event occurrences on it. For instance, if the number of crimes of a particular class increases in some contiguous streets, it is of interest to tell if there is a cluster in that area. Police forces and other public security entities may decide to increase patrolling or to implement other measures to decrease the local crime rate.

Most methods for cluster detection in spatial point data are based on comparison of second order interactions of point processes such as Ripley's $K$-function or the pair correlation function (Ang et al., 2012) and score test statistics (Assunção & Maia, 2007), which require knowledge of the density function of the point process under testing. This is a complicated pace to follow as finding the density function governing the spatial distribution of the points is far from a simple exercise. For example, McSwiggan et al. (2017) propose a density estimator based on diffusions.

Bayesian literature dealing with clustering includes mixture models and random partition methodologies, naturally appearing when a nonparametric approach is undertaking. These have been extended to the spatial setting by incorporating latent variables in the weight structure of the underlying random probability measure (Duan et al., 2007). However, as in the aforementioned approaches, the availability of linear network valued density functions is required to perform clustering.

Here, we present a model that unveils clustering structures on linear networks based on point events. Our proposal induces a spatially dependent random partition model that captures the inherent clustering structure. Specifically, the spatial dependence enters through a penalty function in the corresponding predictive distribution. More importantly, we focus on modeling the occurrence of events on each edge of the linear network instead of modeling the point process itself. This allows us to cluster the edges, preserving the spatial location of the events, but casting aside the complex topology imposed by the linear network. Furthermore, it will be possible to easily estimate the hot-spot locations since they should be related to higher incidence rates.

We also analyze model performance with simulated data on a street network and present an application to real crime incidence data in Mexico City. Although crimes can be classified in different types and severity, we choose armed robberies in a specific zone of the city, due to their high incidence.

## 2 Clustering via random partitions

Common clustering methodologies aim at gathering observations $x_i$, $i = 1, \ldots, n$, into groups. A clustering $\pi$ can be defined as a partition of the set of observations $x = \{x_1, \ldots, x_n\}$ into $k$ nonempty and non-overlapping groups, say $\pi_1, \ldots, \pi_k$, for some $1 \leq k \leq n$, so $\pi = \{\pi_1, \ldots, \pi_k\}$ where $\pi_j \subseteq x$ for all $j$. For the sake of simplicity, partitions will be written as $\pi_1/\cdots/\pi_k$. Observations belonging to the same group are supposed to be more similar among them than any other in a different group. Mathematically, all possible arrangements for $\pi$ is in bijection with the combinatorial class of *set partitions* (Flajolet & Sedgewick, 2009), here denoted by $\mathcal{P}$. Thus, quantifying the uncertainty inherent to a clustering problem can be done by proposing and studying *random partitions*, i.e., $\mathcal{P}$-valued random variables, and their distributions.

A particular class of partition distributions available in the literature comprises the so-called *exchangeable partition probability functions* (EPPFs). This class appears naturally when studying the clustering of exchangeable observations, driven by species sampling processes, and when working with random probability measures (RPMs), the daily-use tool for most Bayesian nonparametric models; see, for example, Hjort et al. (2010) for a thorough review. Any almost surely discrete RPM can be written as

$$\tilde{p}(\cdot) = \sum_{j=1}^{\infty} w_j \delta_{\xi_j}(\cdot), \tag{1}$$

where $\{w_j\}_{j \geq 1}$ and $\{\xi_j\}_{j \geq 1}$ denote independent random sequences of weights and locations, respectively, satisfying $\sum_{i \geq 1} w_i = 1$ almost surely (a.s.), and $\xi_i \sim \nu_0$ independent and identically distributed [iid], with $\nu_0$ a non-atomic distribution. There are several ways to model the sequence of random weights $\{w_j\}_{j>1}$. Perhaps, one of the more often used is the so-called *stick-breaking* representation, which defines them as

$$w_j = v_j \prod_{l<j} (1 - v_l), \tag{2}$$

for $v_1, v_2, \ldots$ a sequence of $(0, 1)$-valued random variables. Some distributions or processes related to this framework are the following: (a) The canonical Dirichlet process with the choice $v_j \sim \text{Be}(1, \theta)$ [iid], for some $\theta > 0$ (Sethuraman, 1994); (b) the two parameter Dirichlet Process when $v_i \sim \text{Be}(1 - \sigma, \theta + i\sigma)$ independent [ind], for $\sigma \in [0, 1)$ with $\theta > -\sigma$ or $\sigma < 0$ with $\theta = m|\sigma|$ and $m \in \mathbb{N}_+$ (Perman et al., 1992); (c) the geometric process with $v_i = \lambda$ and $\lambda$ a $(0, 1)$-valued random variable (Fuentes-García et al., 2010). More general constructions can be found, for example, in Favaro et al. (2016), Gil Leyva Villa et al. (2020), Gil Leyva Villa & Mena (2021); on a different direction, for directly defining random weights $w_j$, see De Blasi et al. (2020).

For the particular case of the Dirichlet process, i.e. when weights are size-biased, the induced EPPF takes the form

$$\Pr(\pi = \pi_1/\cdots/\pi_k) = \rho_0(\#\pi_1, \cdots, \#\pi_k) = \frac{\theta^k}{(\theta)_{n\uparrow}} \prod_{j=1}^{k} \Gamma(\#\pi_j), \tag{3}$$

where $(x)_{n\uparrow} = x(x+1) \cdots (x+n-1)$ is known as the Pochhammer symbol or rising factorial (Ewens, 1972; Antoniak, 1974), and $\#\pi_j$ stands for the size of the $j$th group.

With the above framework in mind, we consider the following model for cluster detection. Let $y_1, \ldots, y_n$ be a dataset to be clustered, and let $\pi$ be a $\mathcal{P}$-valued random partition with prior distribution $\rho_0$. Our interest lies in the posterior distribution of $\pi$, that is

$$p(\pi \mid y_1, \ldots, y_n) \propto \ell(y_1, \ldots, y_n \mid \pi)\rho_0(\pi). \tag{4}$$

Following a model-based approach, the likelihood function $\ell$ is factorized according to the different groups, $\pi_j$, of $\pi$ in such a way that observations belonging to one group are modeled by a single probability distribution $\kappa_j$. It is common to fix such a distribution and only vary its parameter, so $\kappa_j(\cdot) := \kappa(\cdot; \phi_j)$ for some finite dimensional parameter $\phi_j$. Thus, the likelihood function $\ell$ is obtained after marginalizing kernel parameters $\phi_j, j = 1, \ldots, k$, i.e.

$$\ell(y_1, \ldots, y_n \mid \pi) = \prod_{i=1}^{n} \int_{\Phi} \prod_{i \in \pi_j} \kappa(y_i; \phi_j) \nu_0(\mathrm{d}\phi_j),$$

where $\Phi$ represents the support of $\phi_j$. Finally, the clustering model can be written hierarchically as

$$\begin{aligned} y_i \mid \phi, \pi &\sim \kappa(\phi_j)\mathbf{1}(i \in \pi_j) \ [\text{ind}], \quad i = 1, \ldots, n, \\ \phi_j \mid \pi &\sim \nu_0 \ [\text{iid}], \quad j = 1, \ldots, k, \\ \pi &\sim \rho_0, \end{aligned} \qquad (5)$$

where $\nu_0$ is the prior distribution for kernel parameters.

## 3 Clustering over linear networks

As outlined in the Introduction, we are interested in discovering a clustering structure induced by point patterns over linear networks. We define a linear network, $L$, as a geometric simple graph with a finite set of edges $E = \{e'_1, \ldots, e'_m\}$, and where their endpoints form the set of vertices of $L$. Notice that a linear network is embedded in some region $U \subset \mathbb{R}^2$.

Thus, in order to perform clustering over a linear network, we have two sets of measurements: spatial *location* variables, say $v_i \in L$ for $i = 1, \ldots, n$, and their respective responses, $x_i$. These latter could be constant, e.g. $x_i = 1$, meaning that an event occurred at location $v_i$. Each location variable $v_i$ influences the clustering, in the sense that responses close to each other, under some metric, are more likely to be grouped together.

Among the existing literature dealing with spatial clustering, some of them consider location variables $v_1, \ldots, v_n$ as covariates. For example, MacEachern (1999, 2000) defines the dependent Dirichlet process, where an RPM $\tilde{p}$, as in (1), is indexed by some covariate $z$, leading to random densities of the form

$$\tilde{p}_z(\cdot) = \sum_{j=1}^{\infty} w_{j,z} \delta_{\xi_{j,z}}(\cdot).$$

Usually, random atoms are let fixed across the different values of $z$ and only the random weights depend on the covariates. Several generalizations have been developed from here, see, for example, Jo et al. (2017).

On a different approach, Duan et al. (2007) define generalized spatial Dirichlet process models, where the base measure $\nu_0$ for the atoms of (1) is defined over some stationary Gaussian process and the random weights are constructed by means of some multivariate stick-breaking procedure which makes use of the spatial locations. Similarly, Reich & Fuentes (2007) introduce the spatial dependency via a kernel function, depending on the spatial location variables, weighting the random variables generating the sticks $w_j$ in (2). Another way to include these spatial location variables is presented, for example, in Müller et al. (2011) and Page & Quintana (2016). Their approach is of the type of Model (4) where the prior for the partition is a product partition model (Hartigan, 1990), and include an extra term $g$ for each cohesion, which is a function of all covariates associated to the same cluster. On a slightly different approach, Blei & Frazier (2011) modify the predictive distribution for the Dirichlet process, giving spatial dependence to observed clusters, but not to the new ones.

The main hindrance of these approaches for our context is the lack of non-trivial probability distributions over linear networks. Our proposal aims to overcome this difficulty by casting aside the topology induced by the network as follows. We are given a point pattern process over $L$, that is, a set of locations $v_j \in L$, $j = 1, \ldots, m$, indicating the occurrence of some event (e.g. a crime). Let $y_i$ be the random variable defined as the number of events occurred on edge $e'_i$, i.e.

$$y_i = \#\{v_j \in e'_i : j = 1, \ldots, m\}.$$

For our purposes, it is only required to work with the nonzero variables $y_i$; for simplicity we asume $y_i > 0$ for $i = 1, \ldots, n$ for some $n$. Furthermore, we need to define the *new* spatial location variable for $y_i$, say $e_i$; some options are discussed below. With these new variables $y_i$ and $e_i$, our interest is now to cluster the edges $E$ of $L$ through their corresponding $y_i$ using $e_i$ as the spatial location.

Notice that we no longer worry about the complex topology of the linear network, but only on the region $U$ it is contained. Moreover, grouping the network's edges makes sense, since it will allow us to detect the posible hot-spots.

Under this framework, our proposed model for clustering is the following. The base model is the one presented in Equation (4), and detailed in (5), but the spatial information will be incorporated in the prior for the random partition $\pi$, $\rho_0$. Given a cluster $\pi = \pi_1/\cdots/\pi_k$, a location variable $u_j$, $j = 1, \ldots, k$, is introduced and associated to each group $\pi_j$. Thus, counts $y_l$, whose associated *points* $e_l$ are closer to location $u_j$, are more likely to be put together in the corresponding group $\pi_j$. One way to measure the closeness of a point, $\hat{e}$, and a location, $u$, is through a penalty function, for example

$$w(e, u \mid \tau) = \exp\{-\tau(e-u)'(e-u)\}, \qquad (6)$$

for some $\tau > 0$.

Regarding the definition of the new spatial variable $e_i$, it seems appropriate it is a function or statistic of all events recorded along its corresponding edge $e'_i$. We have chosen the centroid for the sake of interpretability. If $v_{i_1}, \ldots, v_{i_m} \in L$ are such that $v_{i_j} \in e'_i$, $j = 1, \ldots, m$ for some $m$, the centroid is defined as

$$e_i = \frac{1}{m} \sum_{j=1}^{m} v_{i_j}.$$

Now it is necessary to incorporate the penalty function $w$ in the partition distribution $\rho_0$. For this purpose, we make use of the joint distribution of the membership variables $d = (d_1, d_2, \ldots, d_n)$. Given a partition $\pi = \pi_1/\cdots/\pi_k$, membership variables $d$ are such that $d_i = j$ if and only if $i \in \pi_j$ for some $1 \leq j \leq k$, and for $i = 1, \ldots, n$. Fuentes-García et al. (2019), Miller (2019) and Gil Leyva Villa & Mena (2021) provide detailed discussions regarding the relationship of these two distributions.

Taking the EPPF induced by the Dirichlet process in Equation (3), the predictive distribution for any $d_i$, $i = 1, \ldots, n$, is

$$\Pr(d_{n+1} = \delta \mid d_1, \ldots, d_n) = \begin{cases} \dfrac{\theta}{n+\theta} & \text{if } \delta \notin \{d_1, \ldots, d_n\}, \\ \dfrac{1}{n+\theta} & \text{if } \delta = d_l \text{ for some } d_l \in \{d_1, \ldots, d_n\}. \end{cases}$$

Therefore, by including the penalty function $w$, we obtained the following distribution.

**Definition 1** *The spatially dependent predictive distribution for the membership variables $(d_1, \ldots, d_n)$, obtained from the EPPF for the Dirichlet process with total mass $\theta > 0$ and penalty function $w$, using Equation (6), is*

$$\Pr(d_i = \delta \mid d_{-i}, e, u, u^*, \tau) \propto \begin{cases} \dfrac{\theta}{n+\theta} w(e_i, u^* \mid \tau) & \text{if } \delta \notin d_{-i}, \\ \dfrac{1}{n+\theta} w(e_i, u_l \mid \tau) & \text{if } \delta = d_l \text{ for some } d_l \in d_{-i}, \end{cases} \qquad (7)$$

where $d_{-i} = \{d_1, \ldots, d_{i-1}, d_{i+1}, \ldots, d_n\}$, for $1 \leq i \leq n$, $e = \{e_1, \ldots, e_n\}$ is the set of spatial variables, $u$ is the set of locations, and $u^*$ is a draw from some non-atomic distribution $\mu_0$ over $U$.

Hence, our proposed model is obtained by extending Model (5) as follows

$$y_i \mid \pi, \phi, e, u \sim \kappa(\phi_j)\mathbf{1}(i \in \pi_j) \quad [\text{ind}], \quad i = 1, \ldots, n, \tag{8}$$
$$\pi \mid e, u \sim \rho_0(e, u),$$
$$\phi_j \mid \pi \sim \nu_0 \quad [\text{iid}], \quad j = 1, \ldots, k,$$
$$u_j \sim \mu_0 \quad [\text{iid}],$$

where $\rho_0(e, u)$ corresponds to the spatial EPPF inherent to Equation (7) in the definition above. The model $\kappa$ for the counts $y_i$ can be any discrete distribution supported over $\{1, 2, \ldots\}$, e.g. the Poisson or negative binomial, and once it is defined, $\nu_0$ will be specified.

Among the different discrete kernel functions $\kappa$, we choose a shifted Poisson of parameter $\lambda$. We say a random variable $Y$ follows a shifted Poisson distribution of parameter $\lambda$, if

$$\Pr(Y = y) = \frac{\lambda^{y-1}e^{-\lambda}}{(y-1)!}, \quad y = 1, 2, \ldots,$$

for some $\lambda > 0$. By using this kernel function, parameter $\lambda$ will contain information regarding the intensity of the occurrence of the recorded events for each detected group.

Completing the elements of Model (8), we have $\phi_j := \lambda_j$, and its prior, $\nu_0$, follows a gamma distribution of parameters $(a, b)$. In addition, $\mu_0$ will be the uniform distribution over $U$. Therefore, the likelihood function takes the form

$$\ell(y \mid \pi, \lambda, e, u) = \prod_{i=1}^{n} \frac{\lambda_j^{y_i-1}e^{-\lambda_j}}{(y_i-1)!}\mathbf{1}(i \in \pi_j) = \prod_{j=1}^{k} \frac{\lambda_j^{\sum_{i \in \pi_j} y_i - n_j}e^{-n_j\lambda_j}}{\prod_{i \in \pi_j}(y_i-1)!},$$

where $n_j = \#\pi_j$ is the size of the $j$th group. The posterior distribution of interest is the following

$$p(\pi, \lambda, u \mid y) \propto \ell(y \mid \pi, \lambda, e, u)p(\lambda \mid \pi)p(\pi \mid e, u)p(u). \tag{9}$$

Due to the complexity of the posterior in (9), it is necessary to resort to numerical methods, specifically we make use of a Gibbs sampler to obtain estimates of the model parameters. At each iteration, it is assumed there are $k$ groups, so the full conditional distribution for each kernel parameter $\lambda_j$, $j = 1, \ldots, k$, is

$$p(\lambda_j \mid \lambda_{-j}, \pi, y) \propto \lambda_j^{\sum_{i \in \pi_j} y_i - n_j + a - 1}e^{-(n_j+b)\lambda_j},$$

which is a gamma distribution of parameters $(\sum_{i \in \pi_j} y_i - n_j + a, n_j + b)$. The second set of parameters corresponds to the locations $u_j$, $j = 1, \ldots, k$, for which full conditional distributions take the form

$$p(u_j \mid u_{-j}, \pi, \tau, e, y) \propto \exp\left\{-\tau n_j u'u + 2\tau u'\left(\sum_{i \in \pi_j} e_i\right)\right\}\mathbf{1}(u_j \in U).$$

Being a bounded distribution, it is straightforward to sample from it.

Sampling for the random partition $\pi$, is done via the membership variables $d_i$, $i = 1, \ldots, n$. For the case there is a new group, $\delta \notin d_{-i}$,

$$\Pr(d_i = \delta \mid d_{-i}, e, \lambda, \lambda^*, u, u^*, \tau, y) \propto \theta w(e_i, u^* \mid \tau)\kappa(y_i, \lambda^*),$$

with $u^*$ and $\lambda^*$ drawn from their respective prior distribution, $\mu_0$ and $\nu_0$. On the other hand, where there is only a switch from one group to another already existing, $\delta = d_l$ for some $l$,

$$\Pr(d_i = \delta \mid d_{-i}, e, \lambda, \lambda^*, u, u^*, \tau, y) \propto w(e_i, u_l \mid \tau)\kappa(y_i, \lambda_l).$$

The total mass parameter $\theta$ can be included in the sampling process as explained, for example, in Escobar & West (1995). Finally, for the penalty function $w$, parameter $\tau$ can be also included in the Gibbs sampler by assigning a gamma prior of parameters $(c, d)$, so its conditional distribution is conjugate and given by

$$p(\tau \mid \pi, u, e, y) \propto \tau^{c-1} \exp\left\{-\tau\left(d + \sum_{i=1}^{n} e'_i e_i - 2 \sum_{j=1}^{k} u'_j \sum_{i \in \pi_j} e_i + \sum_{j=1}^{k} n_j u'_j u_j\right)\right\}.$$

## 4 Simulation study and sensitivity analysis

Our methodology is tested by using two simulated datasets. We are mainly interested in studying the effect of model parameters, which are the penalty parameter $\tau$, the kernel parameter $\lambda$, and the total mass parameter $\theta$.

The first synthetic dataset consists of a sample of 200 event points over a *small* linear network (Figure 1). A sample of size $n = 14$ is obtained after computing the non-zero counts $y_i$; then, their associated centroids $e_i$ are computed. The Gibbs sampler detailed in the previous section was run for 7 000 iterations; posterior estimates were computed using only the last 2 000 of them. A gamma prior of parameters $(1.1, 0.1)$ is set for parameter $\lambda$. The performance of parameters $\theta$ and $\tau$ is studied by assigning them different values. For the total mass parameter $\theta$, its values are chosen such that the prior expected number of groups is 2, 7 and 13. Thus, $\theta$ will take the values 0.3669, 4.8986 and 82.1121. On the other hand, penalty parameter $\tau$ was fixed to $10^r$, for $r = 2, 5, 7, 9$.

The reported estimated clustering corresponds to the posterior modal partition, denoted by $\tilde{\pi}$, a reasonable choice for discrete-valued point estimates. The results of this first simulation study show that the total mass parameter, $\theta$, works as already known for Dirichlet process priors, since it mainly influences the posterior distribution for the number of groups. However, there is not much change in the estimated clustering. Regarding the penalty parameter $\tau$, it can be seen it is of high influence for preserving spatial clustering restrictions. When this parameter is small, our method performs like a traditional clustering technique, since only three groups are detected, corresponding to *small*, *medium* and *large* counts (Figure 2a). On the other hand, when $\tau$ is large, the posterior modal partition correctly incorporates spatial restrictions (Figure 2d). Since all the scenarios tested perform similarly when varying $\theta$, we only present the case $\theta = 4.8986$ in Figure 2; the supplementary material presents the rest of the cases.

### 4.1 Kernel parameter $\lambda$ as a resolution parameter

We now consider a larger simulated dataset in order to illustrate the role of parameter $\lambda$. This second synthetic dataset was obtained by using a bigger linear network and is formed by 522 simulated event points, having $n = 206$ non-zero counts $y_i$ (see Figure 3).

Posterior estimates are obtained from the last 5 000 iterations of the Gibbs sampler, after discarding a first batch of 10 000. For the total mass and penalty parameters, $\theta$ and $\tau$, prior distributions are assigned as follows: a $(1.1, 0.1)$ gamma distribution for $\theta$, and a $(10^{11}, 10^4)$ gamma distribution for $\tau$. The prior for the intensity kernel parameter, $\lambda$, is a $(1.1, 0.1)$ gamma distribution. Figure 4 presents the posterior modal partition. It is worth mentioning that the posterior modal partition has probability 0.002 and contains 58 groups.

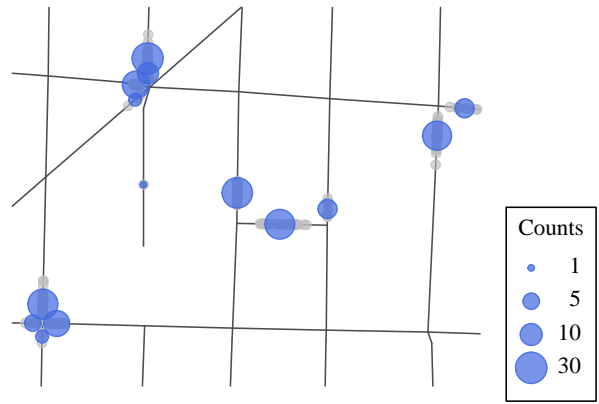

Figure 1: Small simulated dataset. Sampled event points are presented in gray, and the corresponding edge centroid, $e_i$, in blue circles. The size of each circle corresponds to its count value $y_i$.

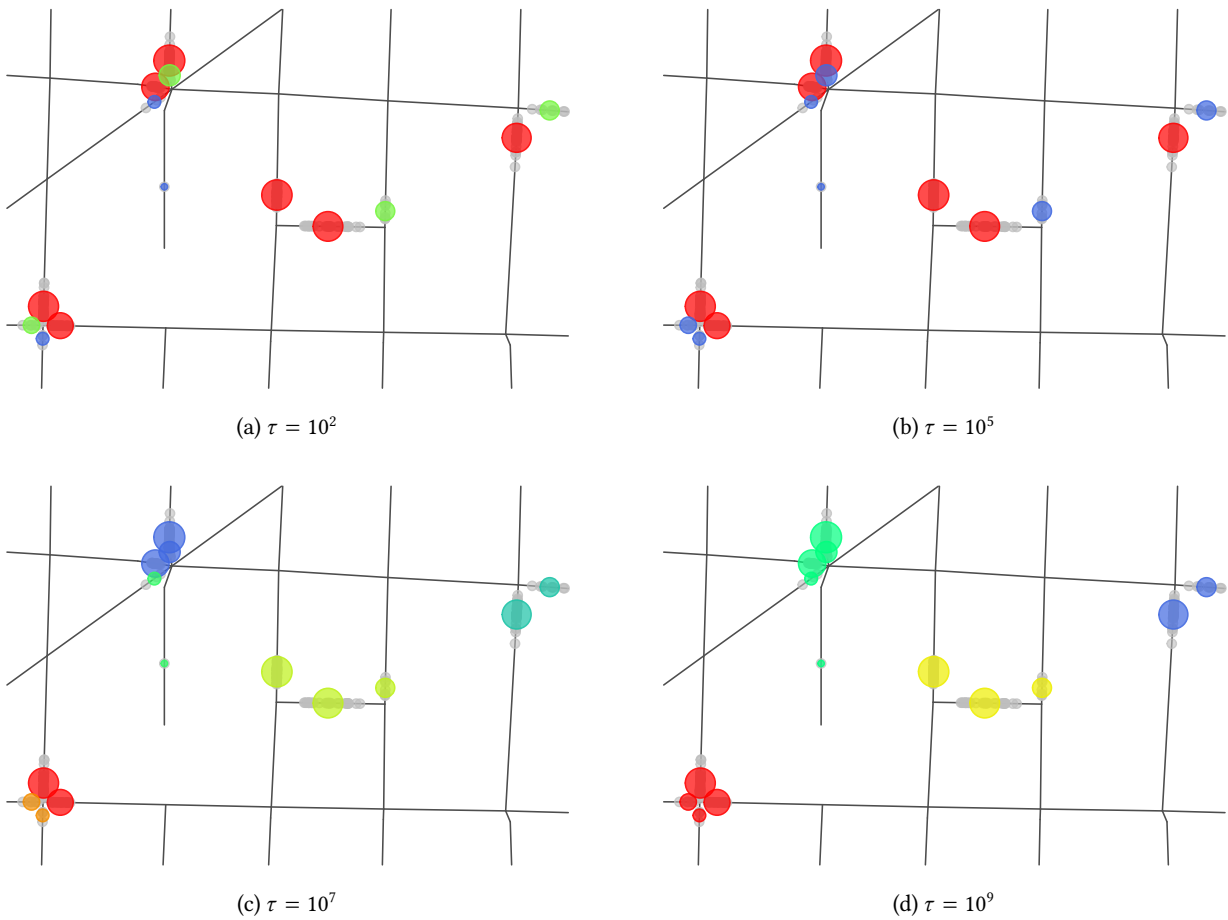

(a) $\tau = 10^2$

(b) $\tau = 10^5$

(c) $\tau = 10^7$

(d) $\tau = 10^9$

Figure 2: Posterior modal partition, $\tilde{\pi}$, for the small simulated dataset, where $\theta = 4.8986$, and $\tau$ varies. Groups are identified by the color of the centroids; colors across panels are totally unrelated.

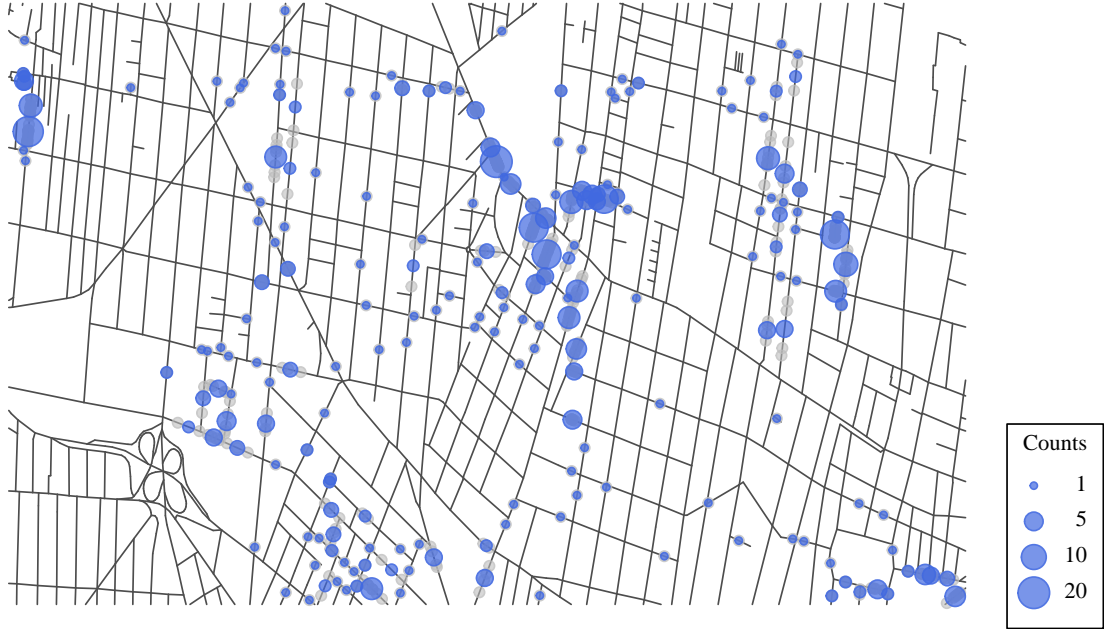

Figure 3: Larger simulated dataset. Sampled event points are presented in gray, and the corresponding edge centroid, $e_i$, in blue circles. The size of each circle corresponds to its count value $y_i$.

Since our main motivation is in identifying hot-spots in a linear network, i.e. set of edges with a particular intensity of events, we explore how the Poisson kernel parameter $\lambda$ can be used to select such relevant clusters.

Given the posterior modal partition $\tilde{\pi}$, for each group $\tilde{\pi}_j$, $j = 1, \ldots, k$, there is a sample, of size $m$, of its corresponding kernel parameter $\lambda_j$, namely $\lambda_{j,l}$ for $l = 1, \ldots, m$ for each $j$. Based on this, we can compute, for example, the mean intensity for each cluster, $\bar{\lambda}_j$, defined as

$$\bar{\lambda}_j := \frac{1}{m} \sum_{l=1}^{m} \lambda_{j,l}, \quad j = 1 \ldots k.$$

Hence, hot-spots can be identified as those clusters having mean intensity above some positive threshold $\lambda^*$, so the resulting restricted clustering will only contain groups $j$ such that $\bar{\lambda}_j \geq \lambda^*$.

Exploring the posterior distribution for kernel parameters $\lambda_j$ conditioned on $\tilde{\pi}$, Figure 5 presents their boxplot. There, we can visually compare the differences among the incidence of events for each group in the modal clustering. Furthermore, Figure 6 displays the selected clusters for a resolution level $\lambda^* \in \{1, 2, 4, 6\}$, with the highest $\lambda^*$ corresponding to higher incidence of events. Moreover, it is also possible to compute the posterior distribution for the number of groups conditioned to this $\lambda^*$. At each iteration, there are some values for the number of groups $k^{(t)}$ together with their corresponding kernel parameter, $\lambda_j^{(t)}$ for $j = 1, \ldots, k^{(t)}$. Then, the posterior distribution for the number of groups, $K_n$, given only clusters with parameter $\lambda$ above $\lambda^*$, is

$$\Pr(K_n = \cdot \mid \lambda^*) \approx \frac{1}{T} \sum_{t=1}^{T} \#\{\lambda_j^{(t)} \geq \lambda^* : j = 1, \ldots, k^{(t)}\},$$

with $T$ the sample size of the Gibbs sampler. Figure 7 presents the posterior distribution for the above values of $\lambda^*$, together with the unconditioned case.

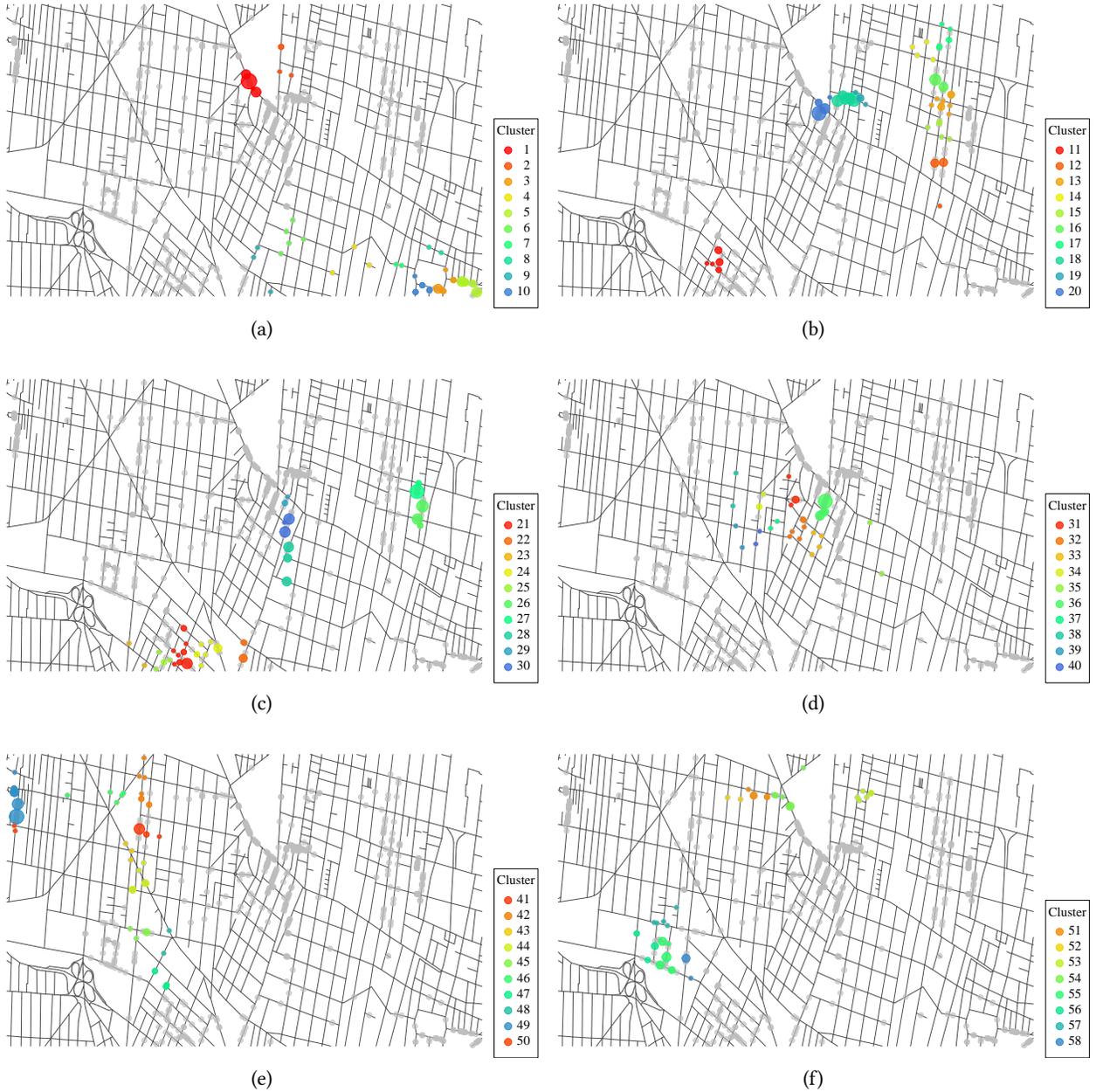

Figure 4: Posterior modal partition for the second simulated dataset. Since there are 58 groups, resulting groups were split into the different panels; each one contains ten groups at most. Colors across panels are completely uncorrelated.

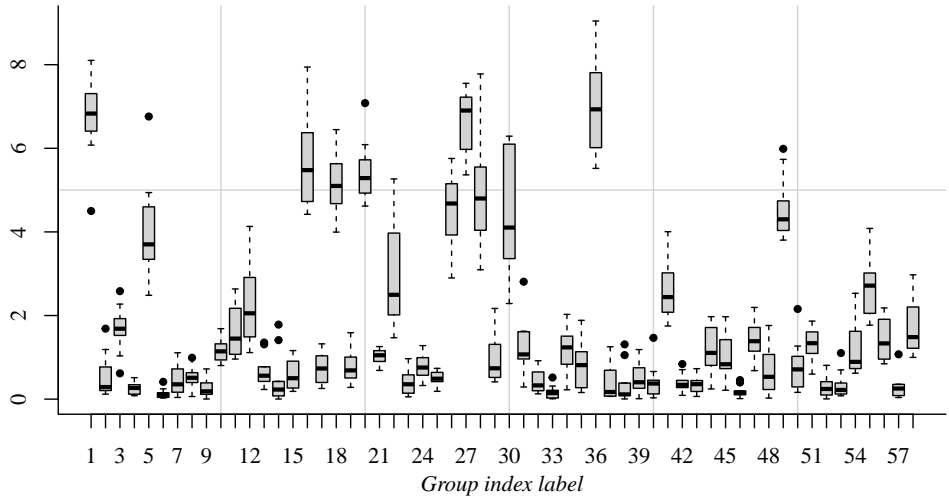

Figure 5: Boxplots for the kernel parameter $\lambda_j$ associated to each group given the posterior modal partition. On the $x$ axis, the label for each group is presented according to those of Figure 4.

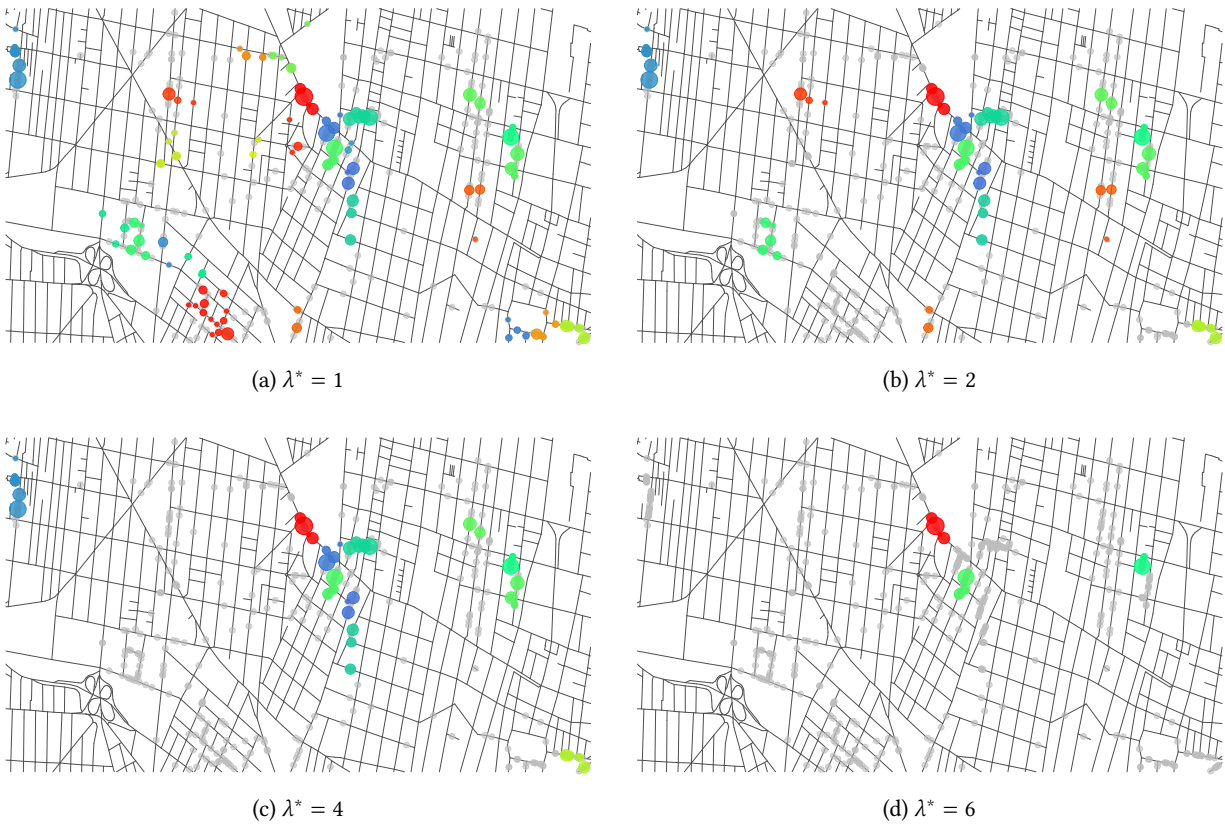

(a) $\lambda^* = 1$

(b) $\lambda^* = 2$

(c) $\lambda^* = 4$

(d) $\lambda^* = 6$

Figure 6: Restricted clustering, based on the modal partition, where posterior mean intensities, $\bar{\lambda}_j$, are above different values of $\lambda^*$: 1, 2, 4 and 6. The size of each circle and its color are as explained in Figures 3 and 4.

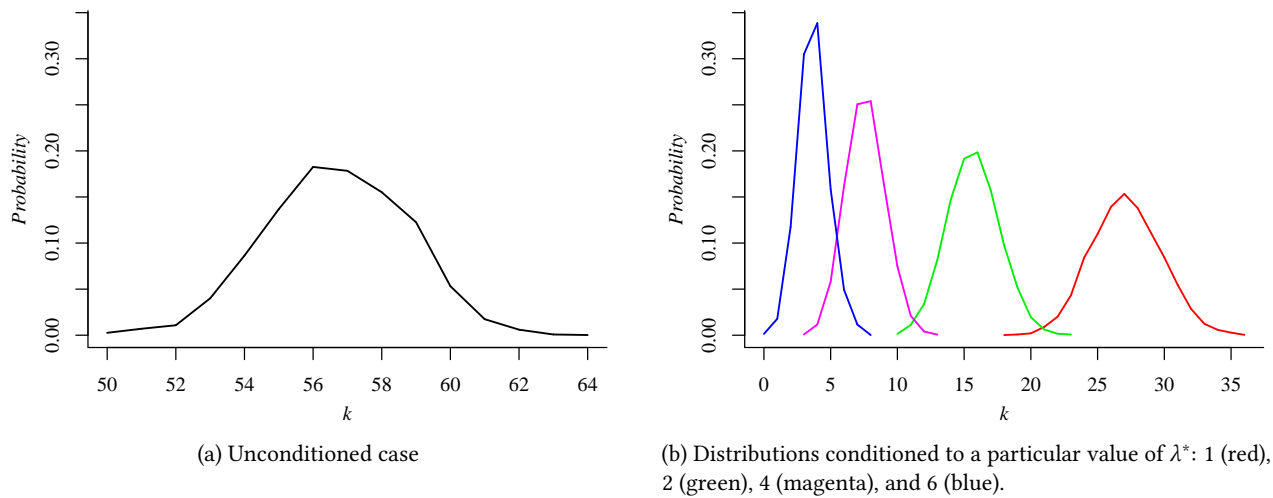

Figure 7: Posterior distributions for the number of groups for the second simulated dataset.

(a) Unconditioned case

(b) Distributions conditioned to a particular value of $\lambda^*$: 1 (red), 2 (green), 4 (magenta), and 6 (blue).

## 5 Application: incidence of violent crimes in Mexico City

The increasing violence and criminality levels in Mexico have become a public safety problem, not only because its effects on the social tissue but also because the material and psychological effects it produces on people (Jiménez Ornelas, 2003). Of particular interest is the incidence of crimes in Mexico City, the capital of the country. Mexico City is the residence of the federal government, and unlike the rest of the states in the country, it has only one police force, under the command of the Secretary of Public Security. Like other cities, Mexico City has areas responding to social factors associated to criminality such as high population density and lack of education and employment, while other areas are associated to factors promoting opportunity for crimes. The reduction of crime incidence comes as a combination of social policies and efficient police actions through intelligence to increase police presence in areas where crime incidence is high. A huge step towards the systematization of crime reports was taken in 2009, when Mexico City Police began recording the geographic location of crimes reported to their force. Despite the high crime incidence, the analysis of crime incidence in Mexico City is difficult. The lack of confidence in the justice system makes that over 80% of crime occurrences go undenounced. According to the civilian organization *México Evalúa*, only 6.8% of crimes are investigated and prosecuted (Piña García & Ramírez-Ramírez, 2019; Mendieta Ramírez, 2019).

Although not all such reports make the way to the justice system and they represent only a small fraction of actual crime incidence, those reports ending in a prosecution by the legal system provide valuable information as they represent a thinned version of the spatial point pattern of actual crime incidence (Valenzuela Aguilera, 2020) A quick police response to a rising crime rate in space and time is only possible if clustering of crime reports is detected promptly. In this section, our clustering method is tested using real data of crime reports.

Data used in this application correspond to cases with an investigation folder, namely those denounced to the authorities by the victim or their legal representatives between January 1st 2018 and December 31st 2019. The database is of public domain and was obtained from the Fiscalía General de Justicia (Attorney's Office) of Mexico City's website[1]. When a crime occurs, the police goes to the crime site to assist the victim, and the location is recorded by the GPS system in the police cars or their mobile phones. Thus, all crime records are georeferenced to a particular location on a street.

The database includes information about time of the crime occurrence, municipality, neighborhood and other information that is considered useful by the authorities and policymakers. It also records many crime types, such as rape, murder and 27 different types of robbery. All these types of robbery were merged into a single crime category for our purposes.

---

[1] https://datos.cdmx.gob.mx/dataset/carpetas-de-investigacion-fgj-de-la-ciudad-de-mexico

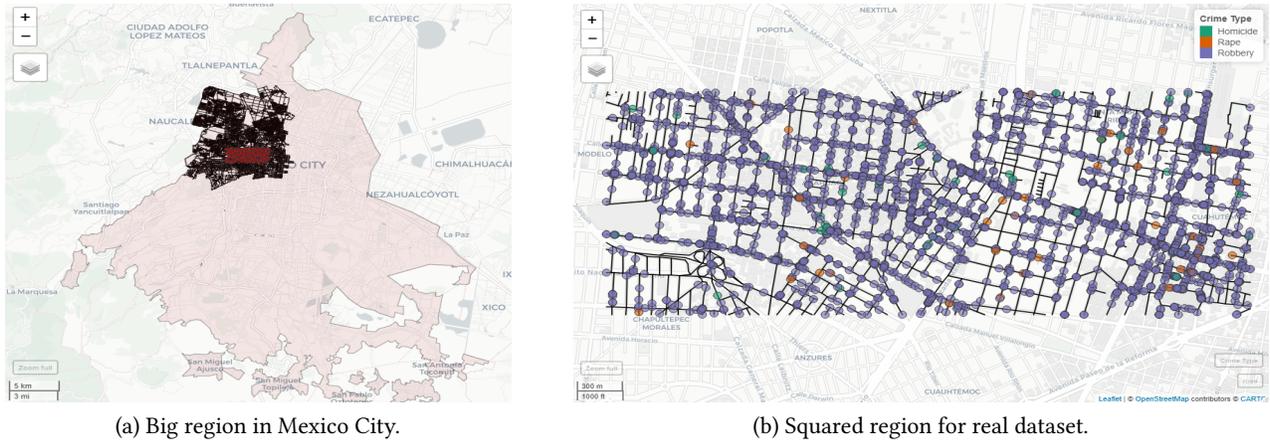

(a) Big region in Mexico City.    (b) Squared region for real dataset.

Figure 8: Region studied in Mexico City. Right panel also contains the distribution of crime events on the road network categorized by crime type.

We selected a section in the northwest of Mexico City as highlighted in black color in Figure 8. The chosen neighborhood contains a mixture of business, industrial as well as wealthy and low income neighborhoods having high crime occurrence. Figure 8a shows in red color the location of a chosen smaller study area within Mexico City. The region highlighted in red is zoomed in Figure 8b with the road network along with locations of individual crime events.

Figure 8b illustrates a total of 2 875 crime incidences distributed as homicide (46), robbery (2 784) and violence rape (45) in the smaller study area. It clearly depicts an uneven distribution of crime types. For further analysis we have selected robbery which has the highest occurrence among the three crime categories. Thus, the rest of the analysis is conducted using 2 784 records of robbery during the entire time period of 2018 and 2019.

Posterior estimates were computed for this dataset using 5 000 iterations after discarding 10 000. Priors for penalty and total mass parameters, $\tau$ and $\theta$, remain as for the second simulated dataset. Different choices were tested for the prior for kernel parameter $\lambda$, from where we chose a $(10, 0.03)$ gamma. The supplementary material contains all the explored cases. We further analyzed the impact of the resolution-level parameter $\lambda^*$ where $\lambda^* \in \{2, 4, 6\}$; see Figure 9, where the dotted horizontal lines in the boxplots depict the three resolution levels. Additionally, Figure 10 depicts the locations of robbery clusters on the street network for two resolution levels $\lambda^*$, 4 and 6, which can be understood as the more relevant hot-spots. In the same figure, clusters are shown along with road segments highlighted in red.

Despite the high crime incidence in the selected subarea of Mexico City, the proposed method is able to detect the presence of several clusters in the zone. All these clusters take place in areas where a mix of small stores, offices and metro or train stations are located. Only the cluster at the center of the analyzed region is located in a poor residential neighborhood. It is not clear if the areas where clusters occur are gang territories or not, but they are clearly zones where the mix of different economic activities attracts many potential victims.

Our proposal has the advantage that the dynamics of crime clustering may be detected in a relatively fast and simple way. Although the convergence speed of the Gibbs sampler depends on the number of edges, if super-computing is not available, i.e., one has limited computational power, our method can be implemented in targeted parts of a city where local authorities need prompt results to trigger police assistance. Another potential use of this study could be the forecasting of true crime incidence rates. Indeed, as recently reported in the National Survey of Victimization and Perception of Public Safety 2020, there is a significantly percentage of crimes not reported. See `inegi.org.mx/programas/envipe/2020` for further details.

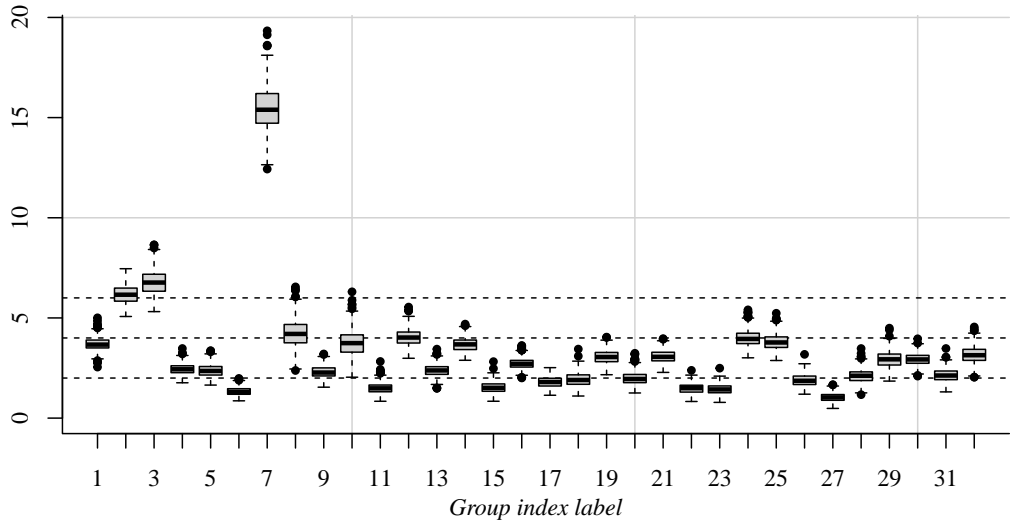

Figure 9: Boxplots for the kernel parameter $\lambda_j$ associated to each group given the posterior modal partition. On the $x$ axis, the label for each group is presented. Horizontal dotted lines correspond to the different values for $\lambda^* \in \{2, 4, 6\}$.

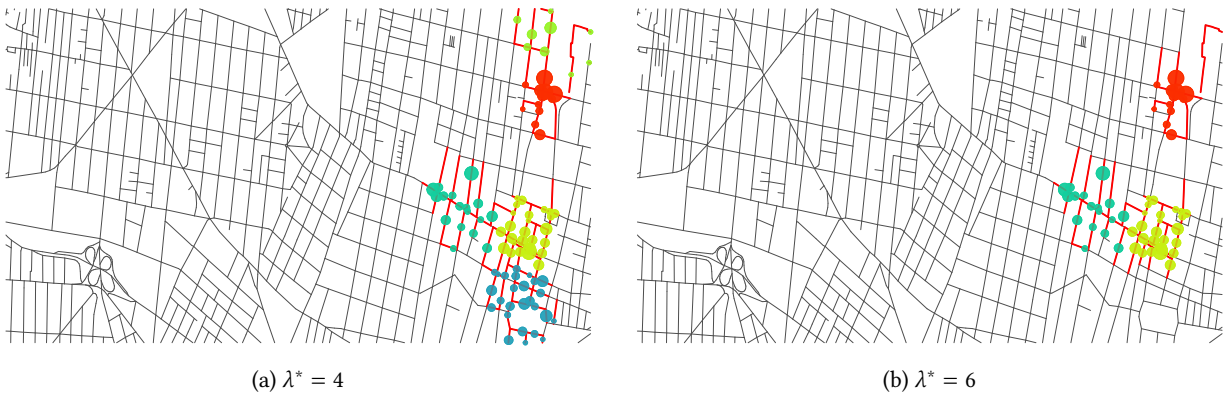

(a) $\lambda^* = 4$  (b) $\lambda^* = 6$

Figure 10: Road segments along with clusters for the street network; cases $\lambda^* \in \{4, 6\}$.

# 6 Concluding remarks

We have presented a simple yet effective model for clustering constrained on linear networks based on point events using random partitions under a nonparametric Bayesian approach. The proposal of removing the topology induced by the linear network, and modeling the occurrence of events over each edge instead, greatly simplifies the clustering task. Furthermore, making the random partition distribution spatially dependent through the penalization function $w$ allows us to reveal clusters of high incidence of events. In our application, those events are armed robberies, of particular interest to citizens and law forces. The Poisson kernel parameter $\lambda$ included in our model controls the resolution of high event incidence, which in the application at issue, helps to identify street configurations with high crime record. All this said, our methodology could be used for other applications as described in Chapter 17 of Baddeley et al. (2016).

We believe our proposal adds a valuable tool to the existing clustering techniques over spatial point patterns. The key fact of the network structure makes this proposal new in this field, and can be considered an attractive while easy-to-use tool in the analysis of point patterns over linear networks.

Here, we have used the clustering induced by the modified Dirichlet process, however other nonparametric priors, such as those belonging to the Gibbs-type family, could also be used.

# Acknowledgement

A.F. Martínez, C. Díaz-Avalos and R.H. Mena thankfully acknowledge the support of PAPIIT project number IG100221. J. Mateu was partially supported by project PID2019-107392RB-I00/AEI/10.13039/501100011033.

# Supplementary material

For the small synthetic dataset, several sampling specifications were run; similarly, for the real data application, different prior specifications were used. Posterior estimates are presented for all these cases.

# Additional simulations for clustering constrained on linear networks

Asael Fabian Martínez

Departamento de Matemáticas, Universidad Autónoma Metropolitana, Unidad Iztapalapa, Ciudad de México, Mexico.

`fabian@xanum.uam.mx`

Somnath Chaudhuri

Research Group on Statistics, Econometrics and Health (GRECS), University of Girona, Spain, and CIBER of Epidemiology and Public Health (CIBERESP), Spain.

`chaudhuri.somnath@udg.edu`

Carlos Díaz-Avalos

IIMAS, Universidad Nacional Autónoma de México, Ciudad Universitaria, Mexico.

`carlos@sigma.iimas.unam.mx`

Pablo Juan

IMAC, University Jaume I, Castellón, Spain.

`juan@uji.es`

Jorge Mateu

Department of Mathematics, University of Jaume I, Castellón, Spain.

`mateu@uji.es`

Ramsés H. Mena

IIMAS, Universidad Nacional Autónoma de México, Ciudad Universitaria, Mexico.

`ramses@sigma.iimas.unam.mx`

## 1 Simulated dataset

Figures 1 and 2 display the posterior modal partition for the small dataset, where total mass parameter varies. Figure 3 presents the posterior distribution for the number of groups.

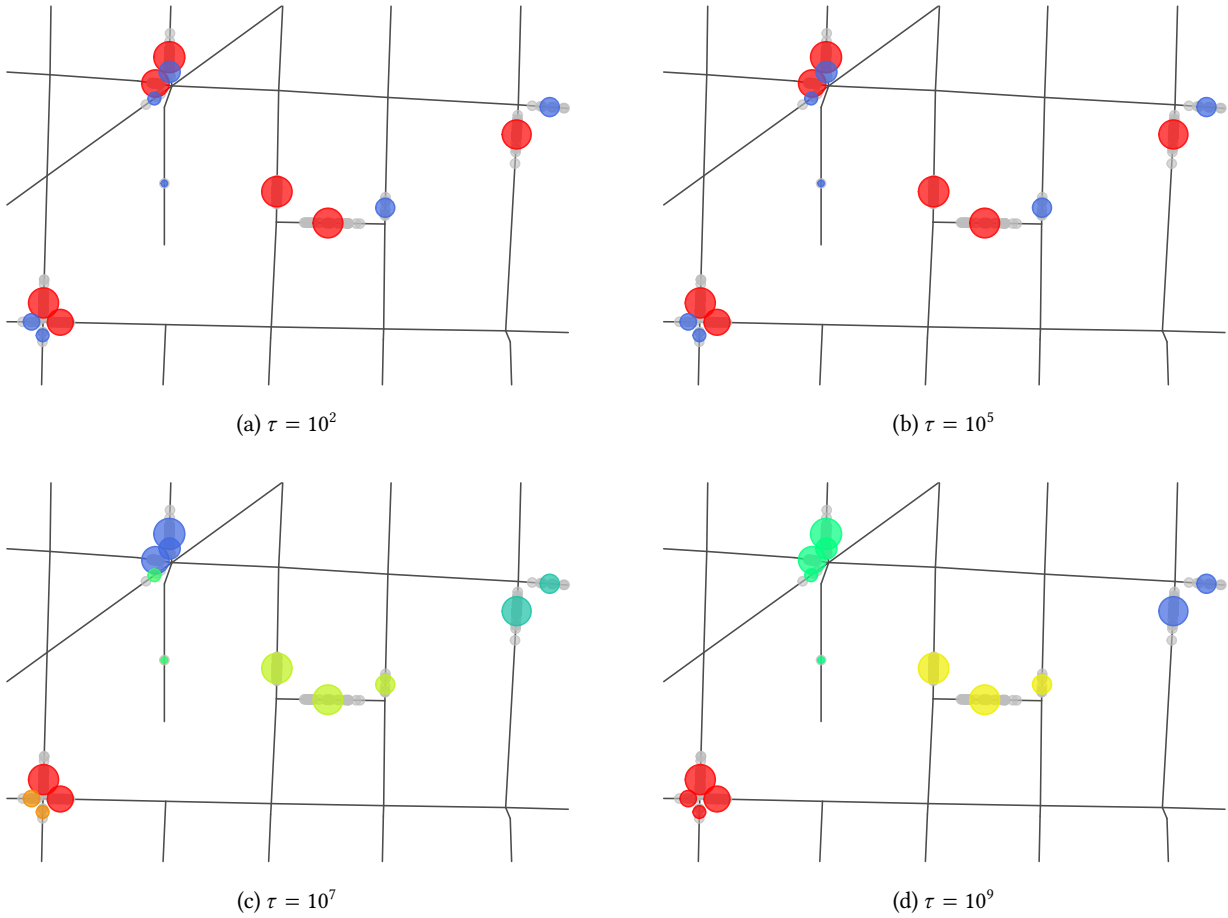

Figure 1: Posterior modal partition for the small simulated dataset, where $\theta = 0.3669$, and $\tau$ varies. Groups are identified by the color of the centroids; colors across panels are totally unrelated.

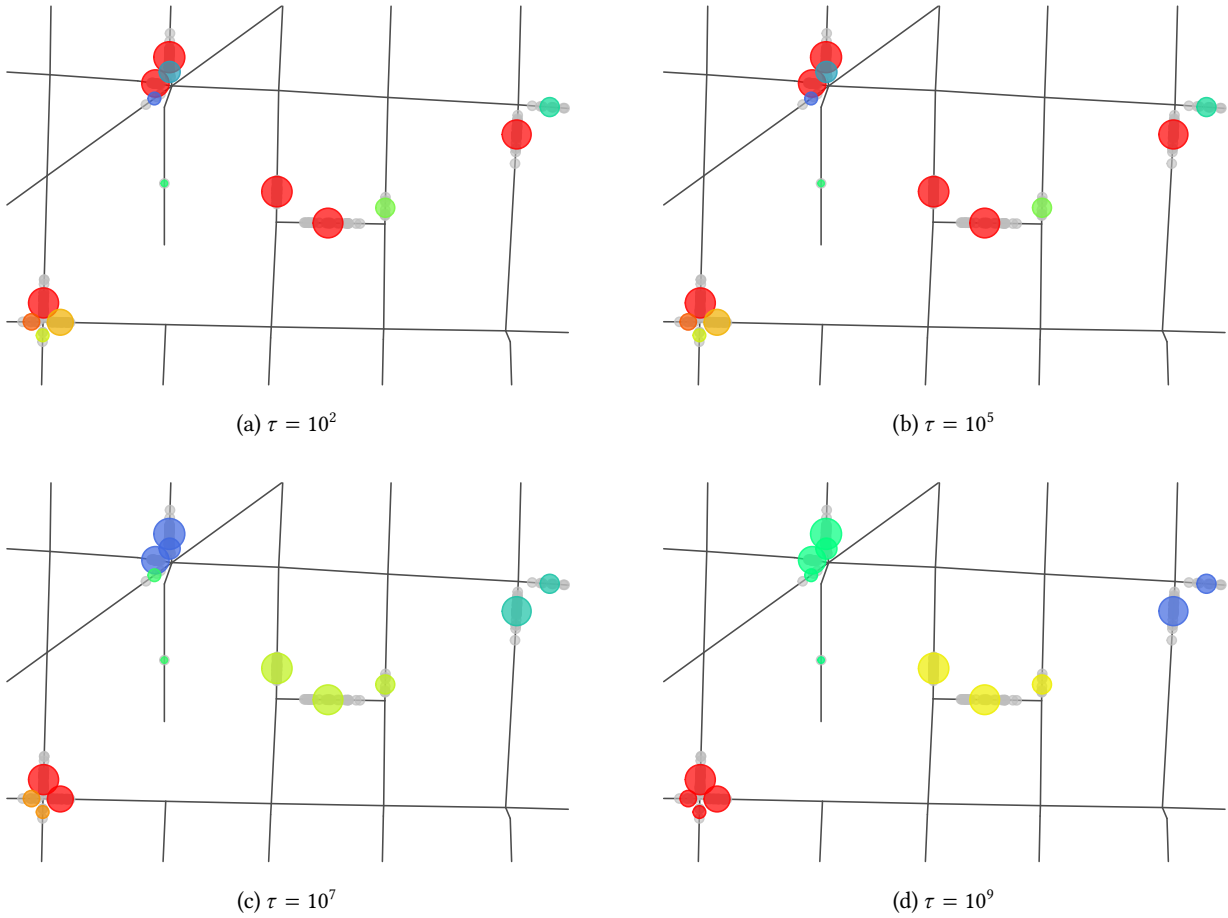

(a) $\tau = 10^2$

(b) $\tau = 10^5$

(c) $\tau = 10^7$

(d) $\tau = 10^9$

Figure 2: Posterior modal partition for the small simulated dataset, where $\theta = 82.1121$, and $\tau$ varies. Groups are identified by the color of the centroids; colors across panels are totally unrelated.

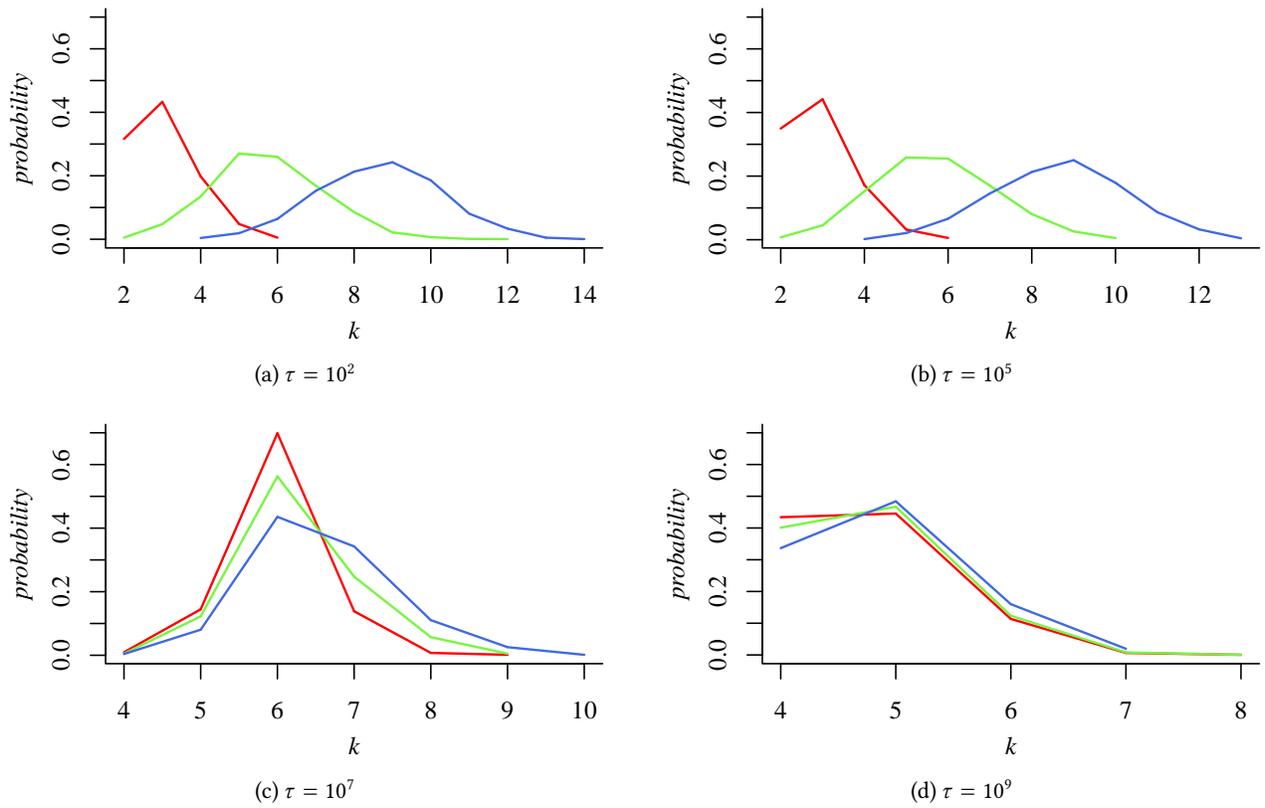

Figure 3: Posterior distribution for the number of groups for the small simulated dataset and different values of $\theta$: 0.3669 (red), 4.8986 (green), and 82.1121 (blue).

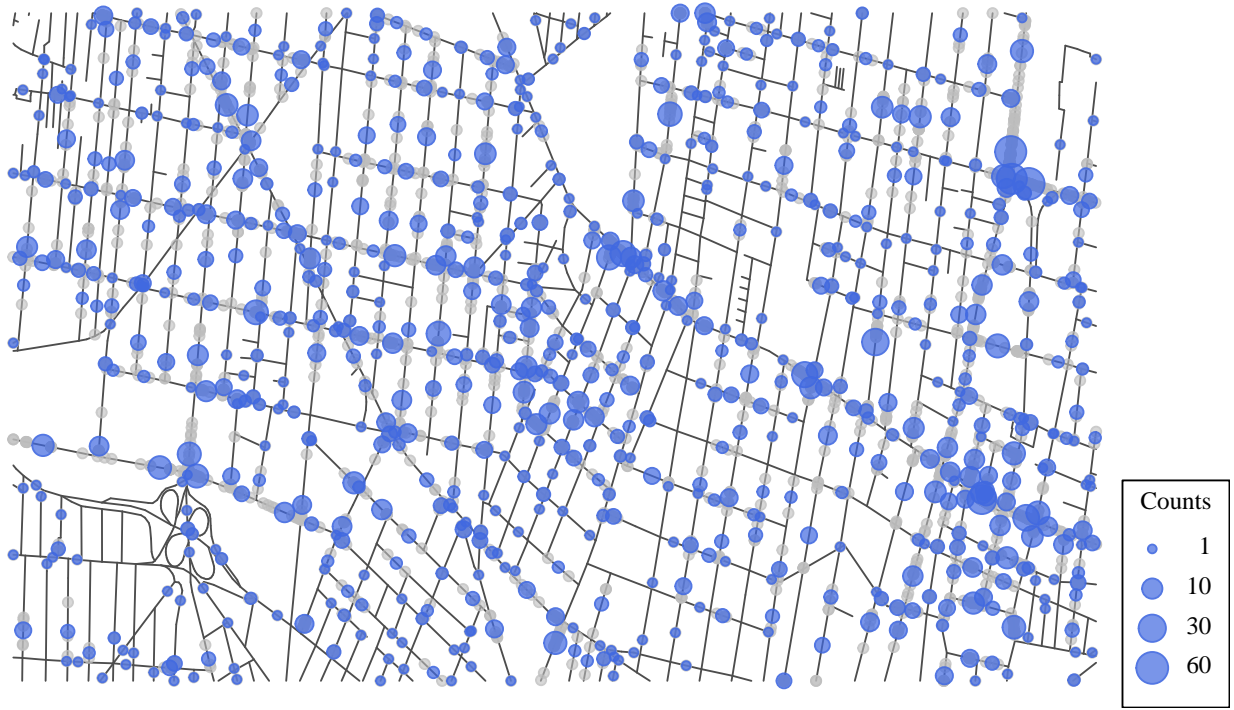

Figure 4: Real dataset

| Model | (a, b) | Model | (a, b) |
|-------|--------|-------|--------|
| A | (4, 0.1) | E | (5, 0.05) |
| B | (10, 0.01) | F | (15, 0.03) |
| C | (10, 0.05) | G | (15, 0.05) |
| D | (10, 0.10) | H | (15, 0.10) |

Table 1: Hyper-parameters used for the Poisson kernel parameter $\lambda$ gamma prior for the real dataset.

## 2 Real dataset

For the real dataset, different prior distributions were tested for the Poisson kernel parameter $\lambda$; Table 1 displays the cases tested. In all of them, the MCMC was run for 10 000 burn-in iterations and a sample of size 5 000 was taken afterwards. The prior for parameter $\tau$ was fixed as a gamma distribution of parameters $(10^{11}, 10^4)$; similarly for the total mass parameter $\theta$, a gamma distribution of parameters $(1.1, 0.1)$ was fixed.

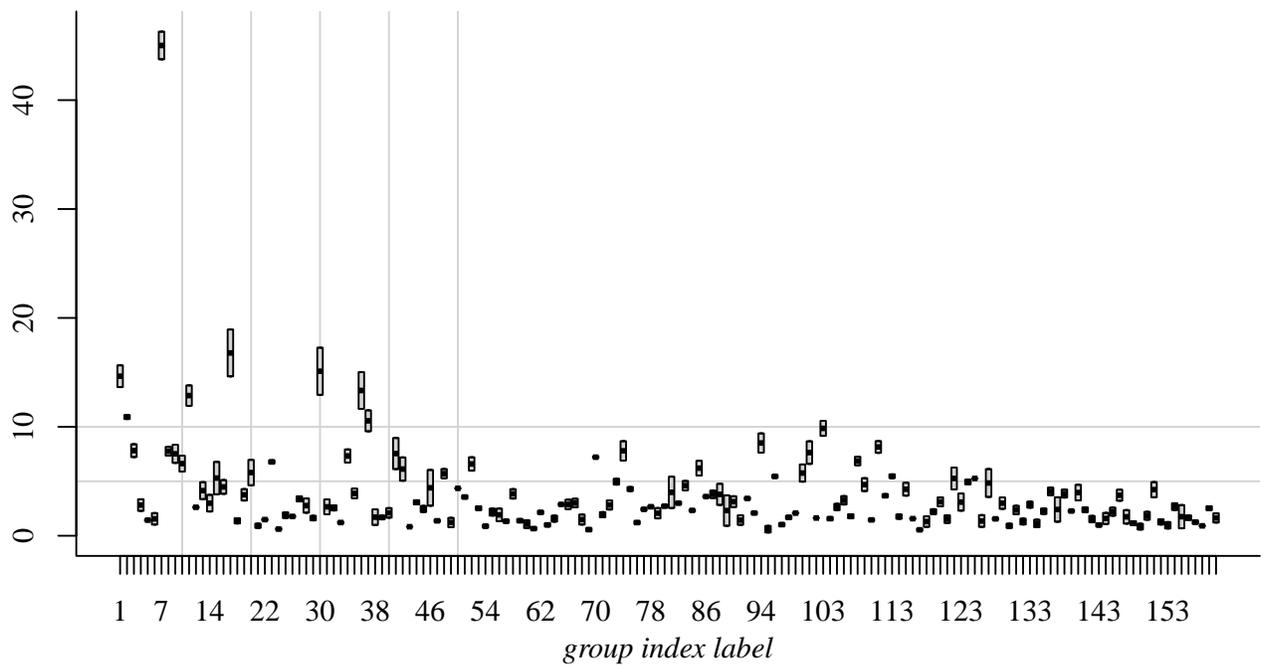

Figure 5: Boxplot for kernel parameters $\lambda_j$ associated to each group given the posterior modal partition; Model A.

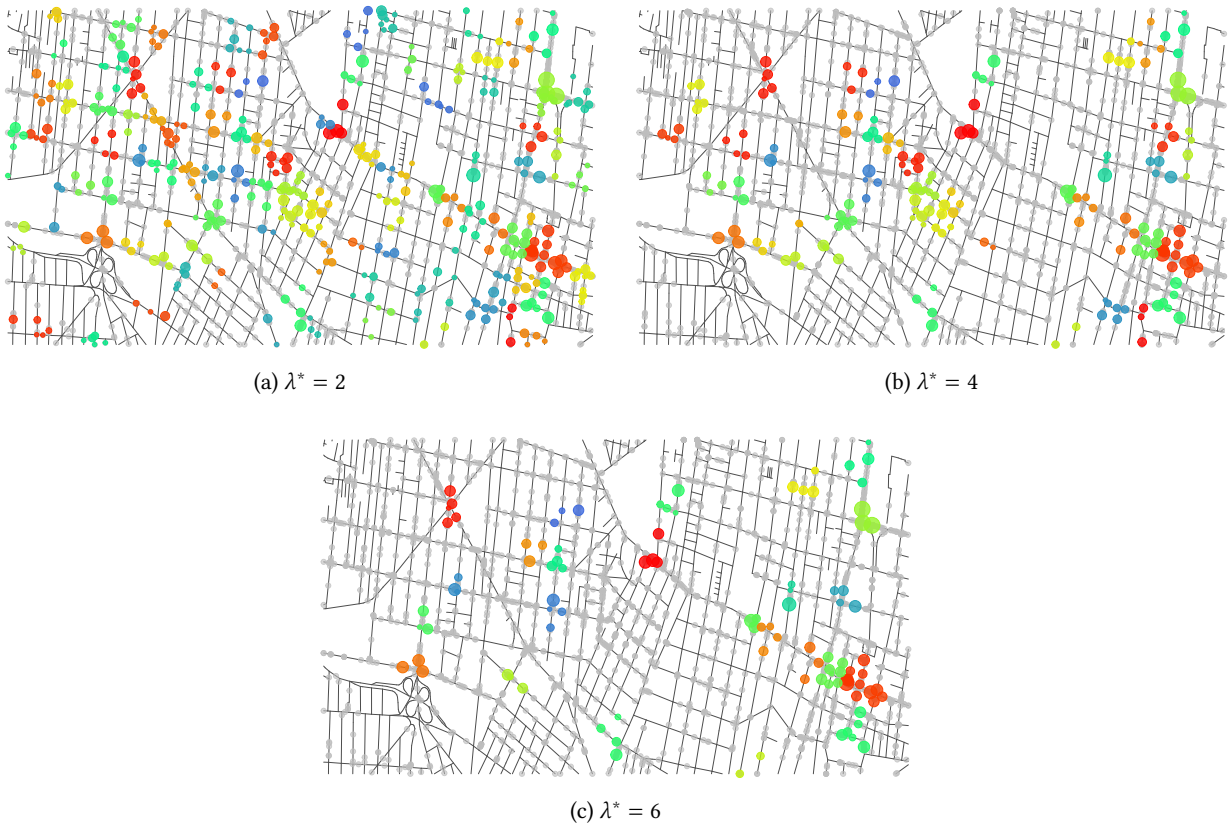

Figure 6: Restricted clustering, based on the modal partition, where posterior mean intensities, $\bar{\lambda}_j$, are above different values of $\lambda^*$: 2, 4 and 6; Model A.

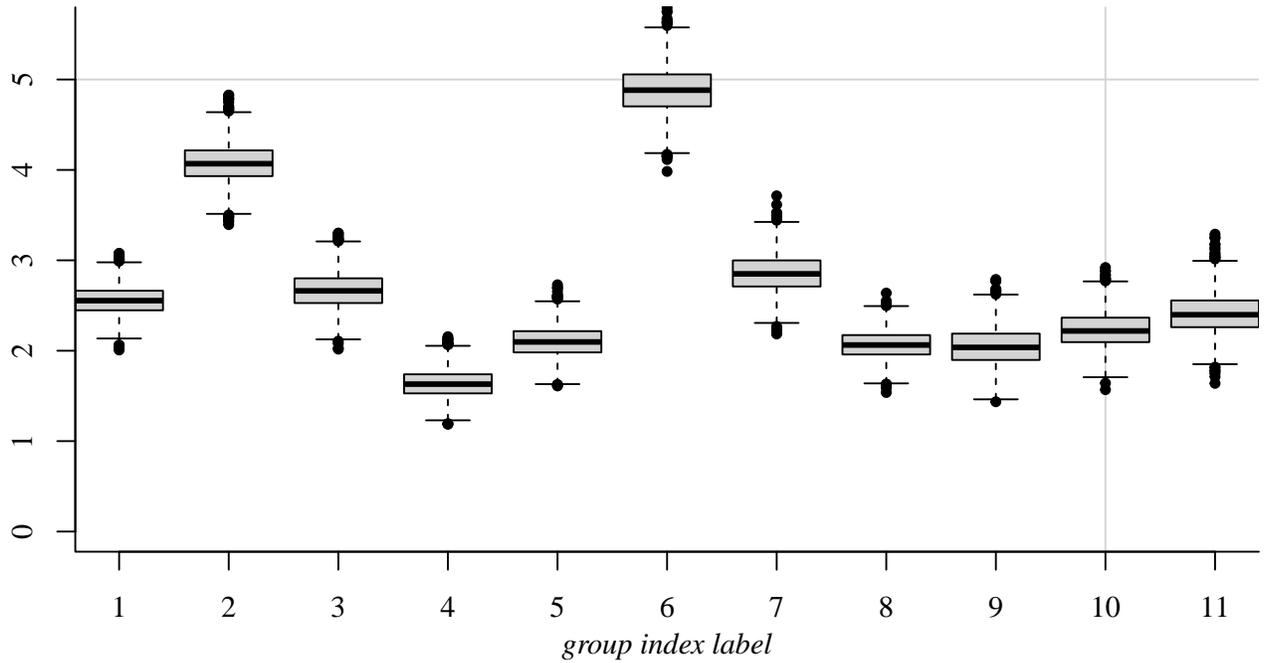

Figure 7: Boxplot for kernel parameters $\lambda_j$ associated to each group given the posterior modal partition; Model B.

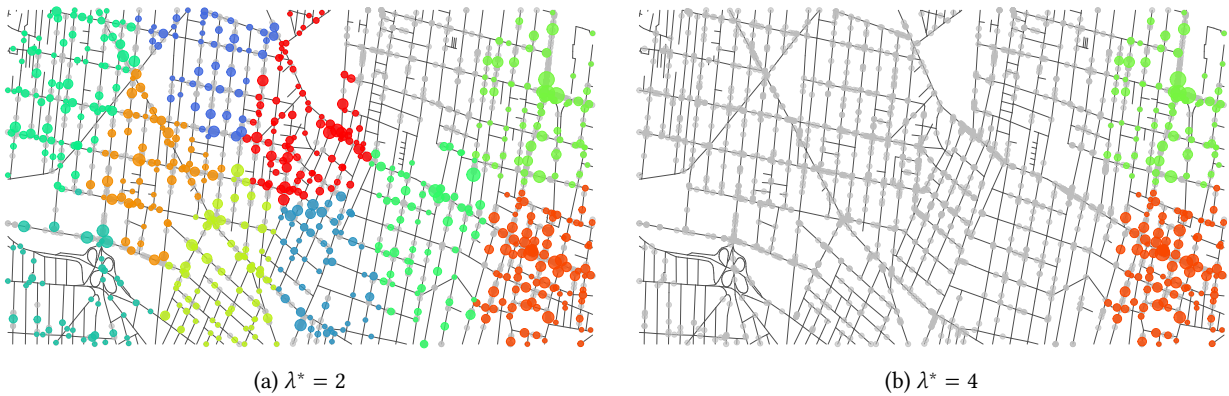

(a) $\lambda^* = 2$

(b) $\lambda^* = 4$

Figure 8: Restricted clustering, based on the modal partition, where posterior mean intensities, $\bar{\lambda}_j$, are above different values of $\lambda^*$: 2, 4 and 6; Model B.

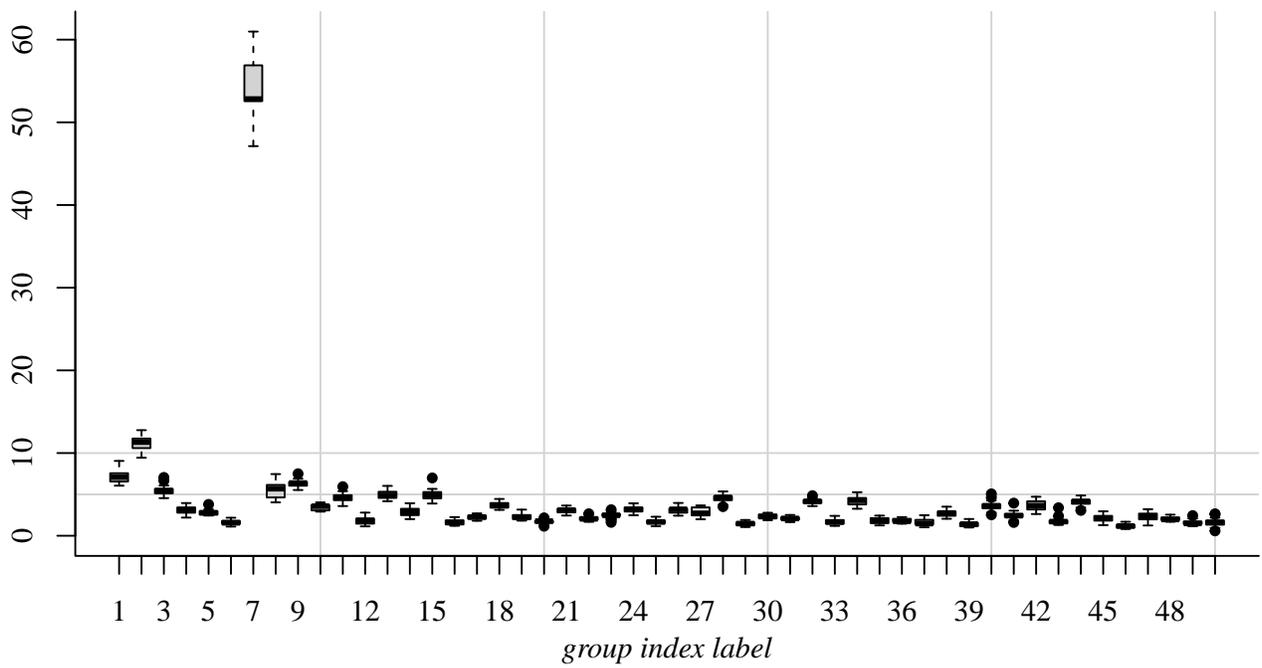

Figure 9: Boxplot for kernel parameters $\lambda_j$ associated to each group given the posterior modal partition; Model C.

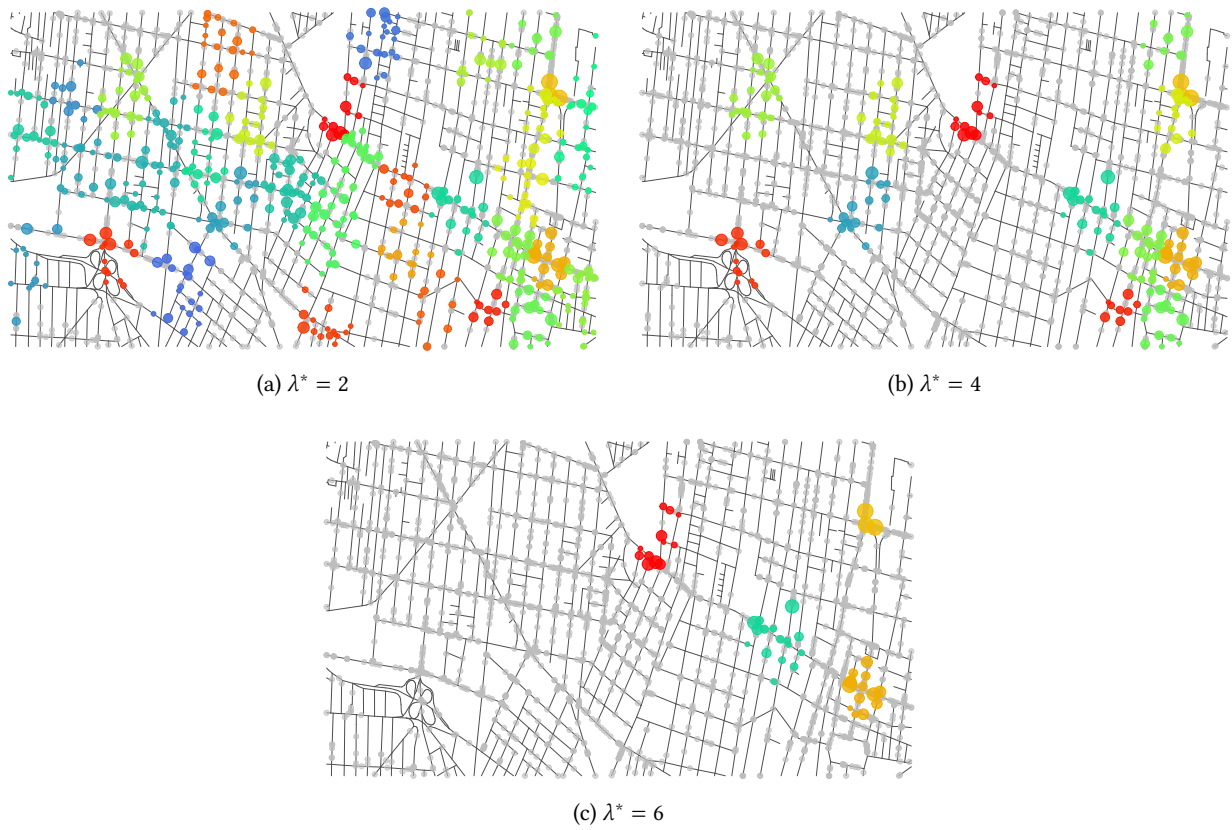

Figure 10: Restricted clustering, based on the modal partition, where posterior mean intensities, $\bar{\lambda}_j$, are above different values of $\lambda^*$: 2, 4 and 6; Model C.

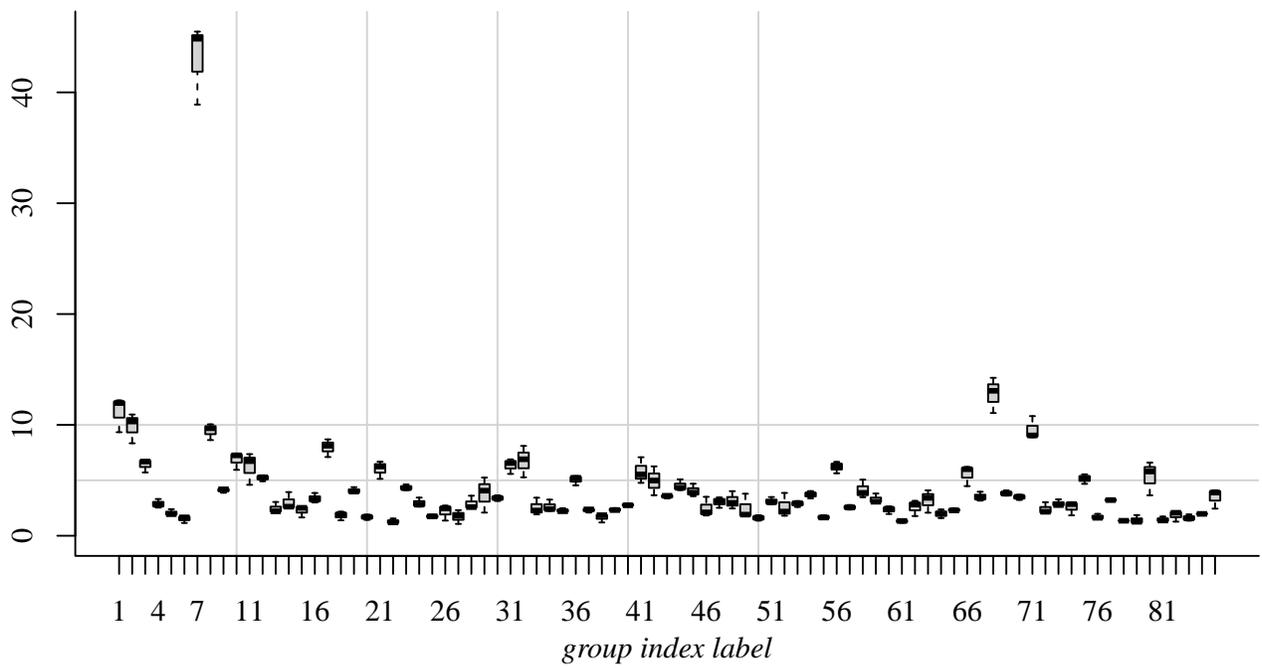

Figure 11: Boxplot for kernel parameters $\lambda_j$ associated to each group given the posterior modal partition; Model D.

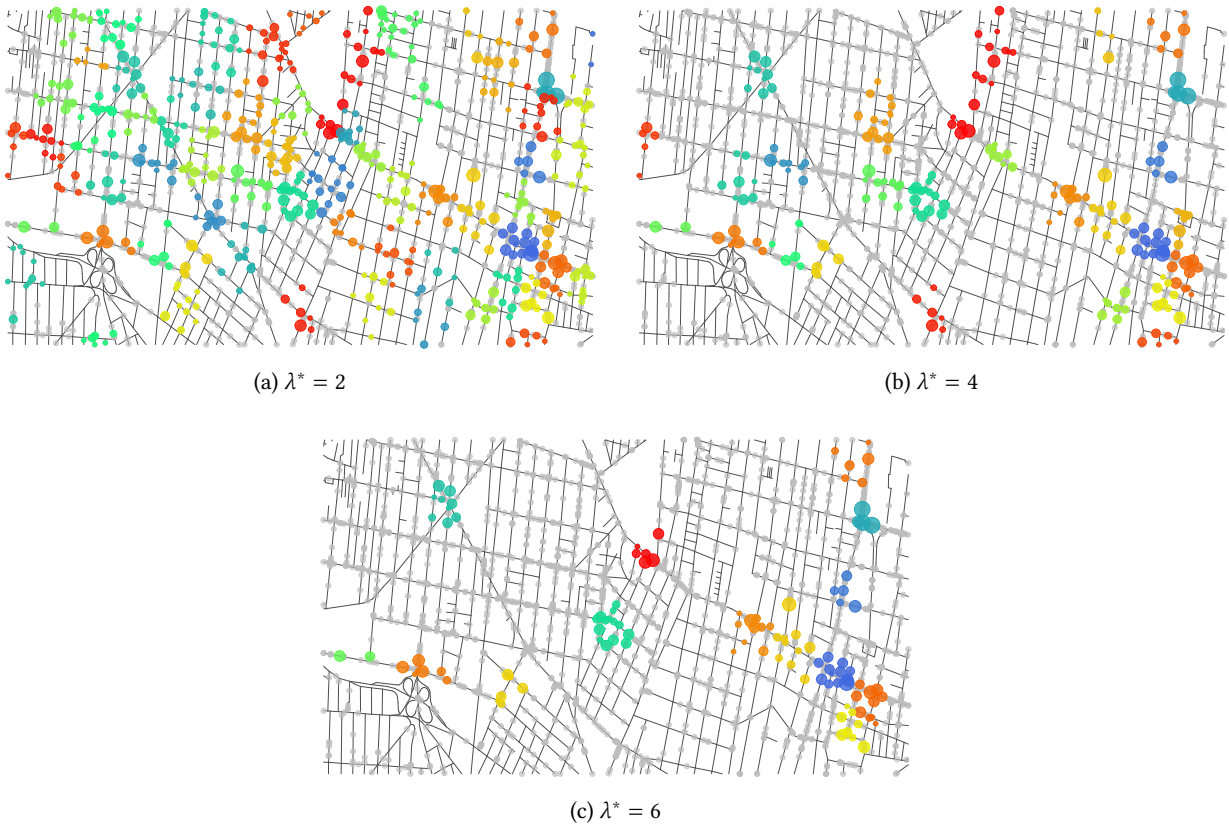

(a) $\lambda^* = 2$

(b) $\lambda^* = 4$

(c) $\lambda^* = 6$

Figure 12: Restricted clustering, based on the modal partition, where posterior mean intensities, $\bar{\lambda}_j$, are above different values of $\lambda^*$: 2, 4 and 6; Model D.

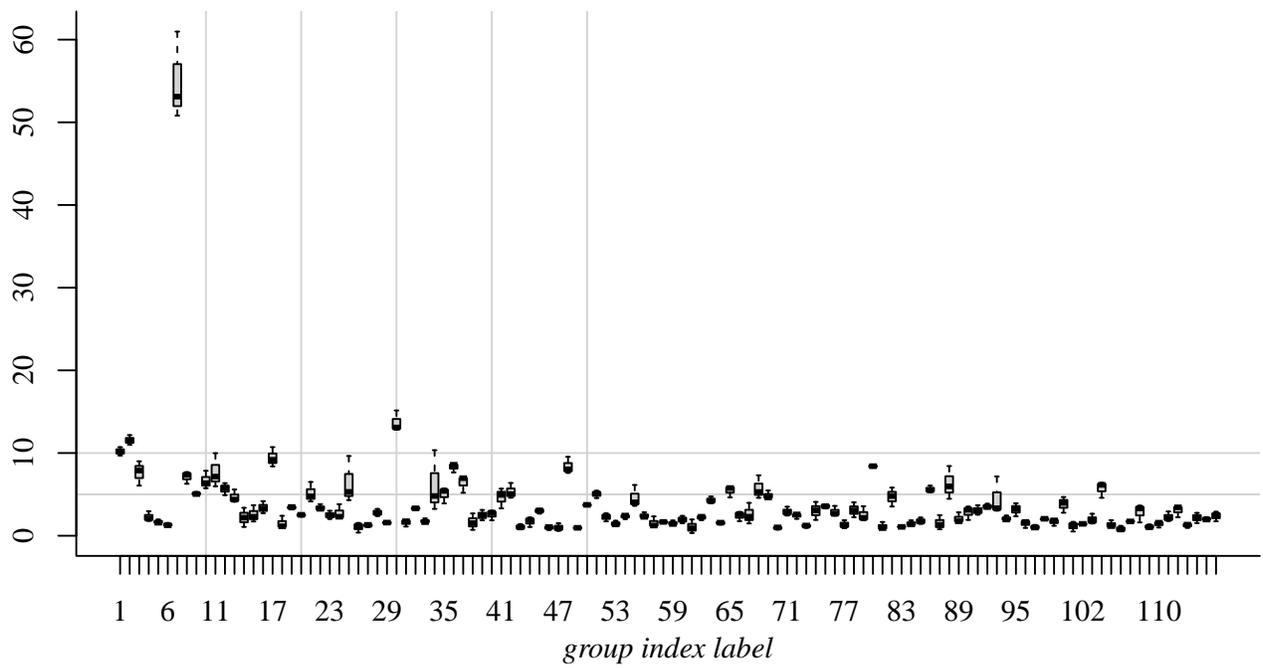

Figure 13: Boxplot for kernel parameters $\lambda_j$ associated to each group given the posterior modal partition; Model E.

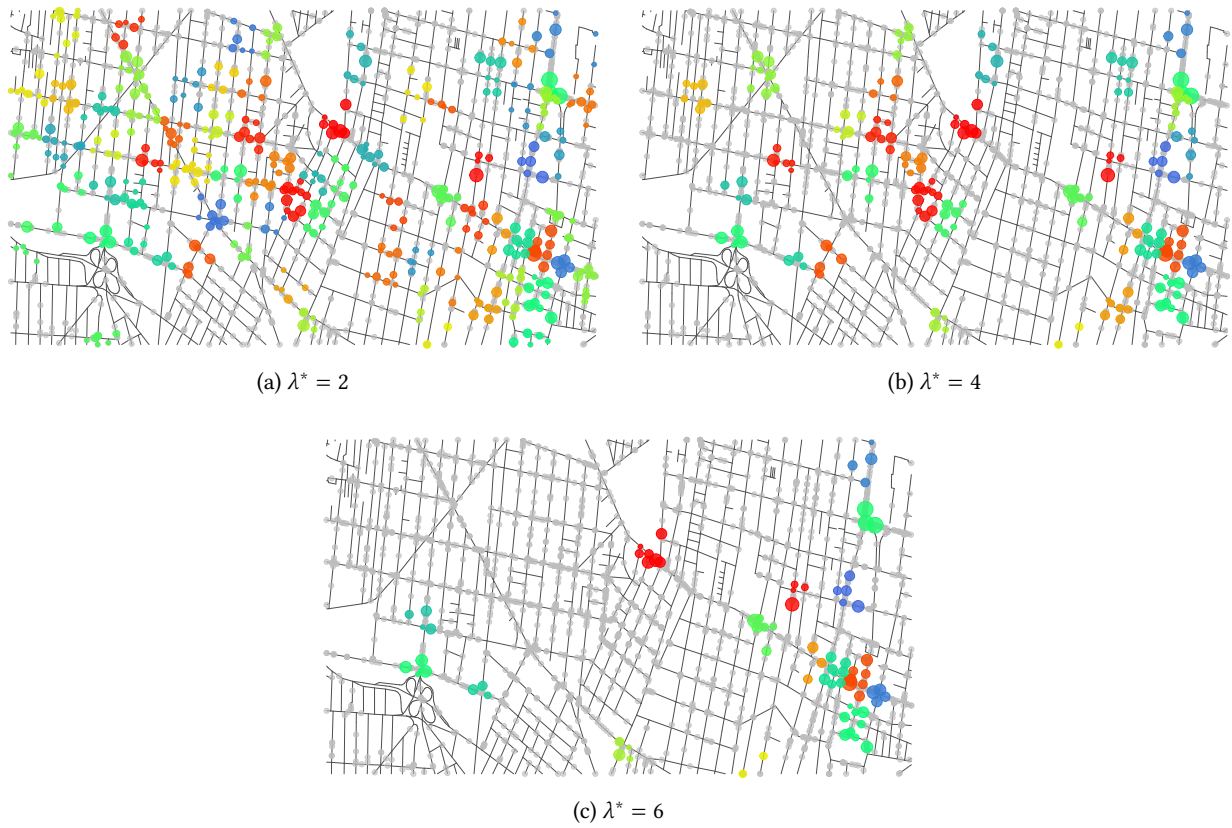

Figure 14: Restricted clustering, based on the modal partition, where posterior mean intensities, $\bar{\lambda}_j$, are above different values of $\lambda^*$: 2, 4 and 6; Model E.

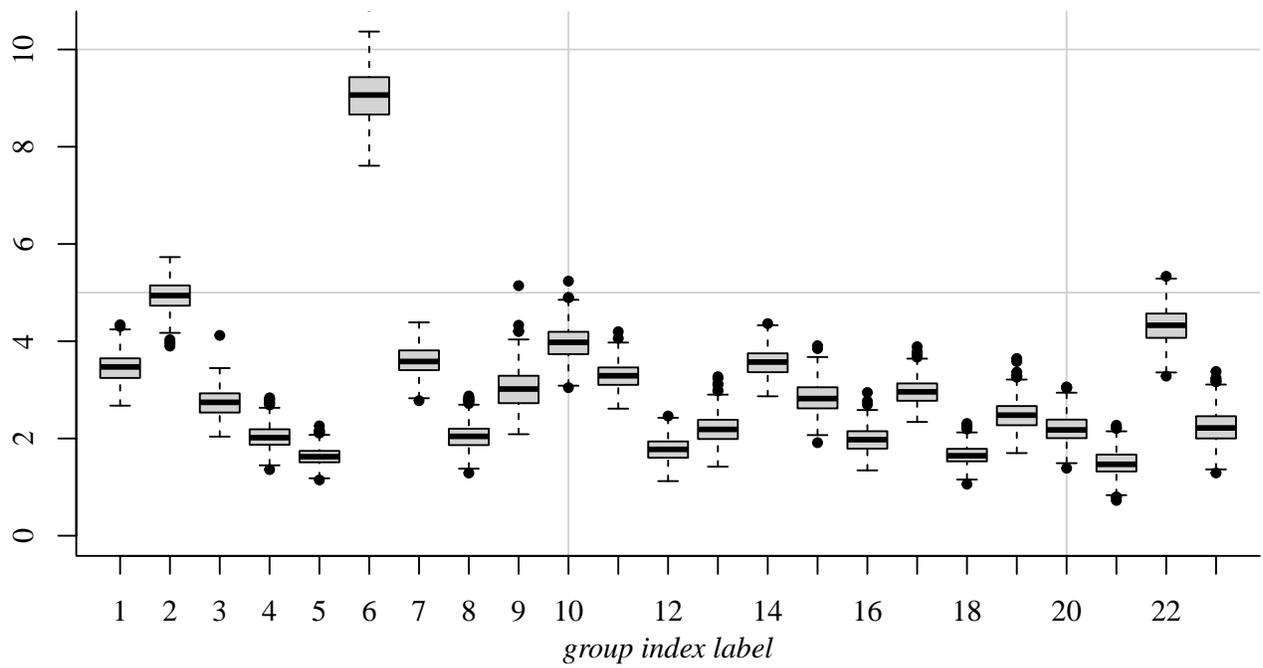

Figure 15: Boxplot for kernel parameters $\lambda_j$ associated to each group given the posterior modal partition; Model F.

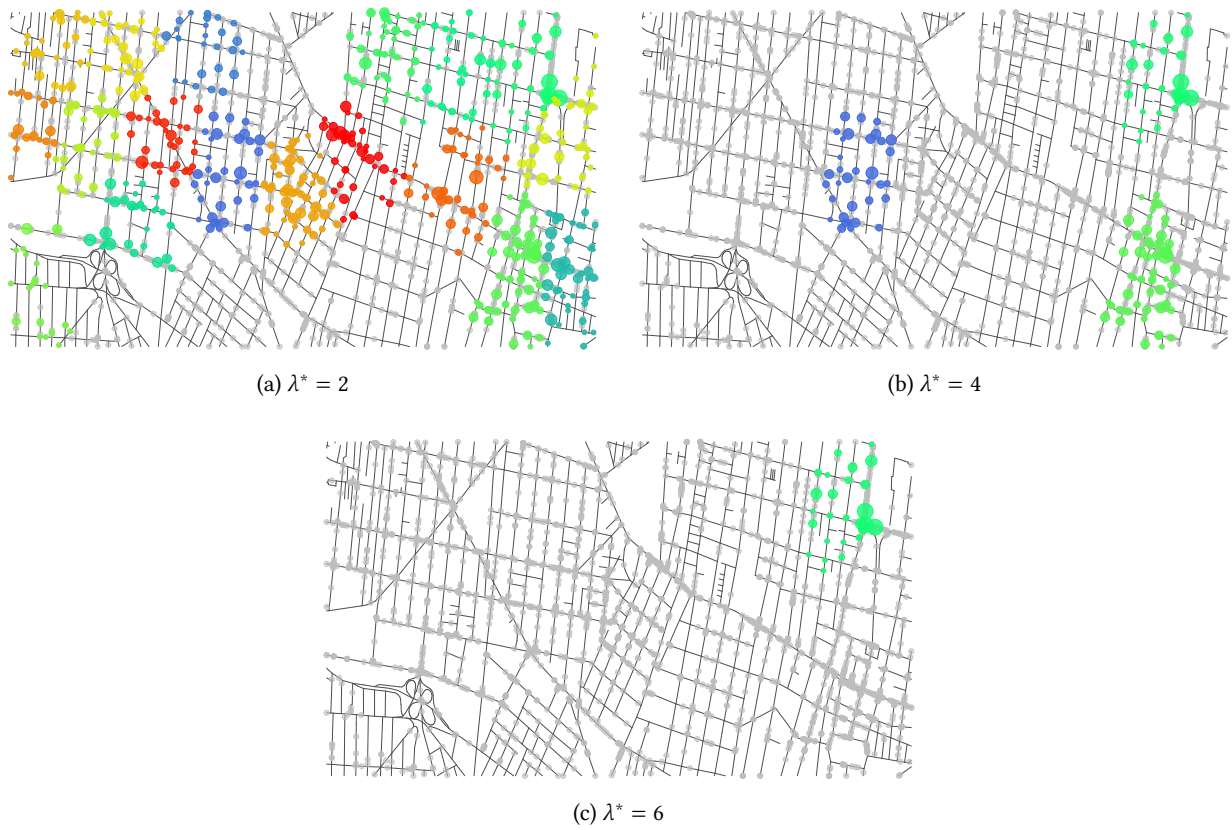

Figure 16: Restricted clustering, based on the modal partition, where posterior mean intensities, $\bar{\lambda}_j$, are above different values of $\lambda^*$: 2, 4 and 6; Model F.

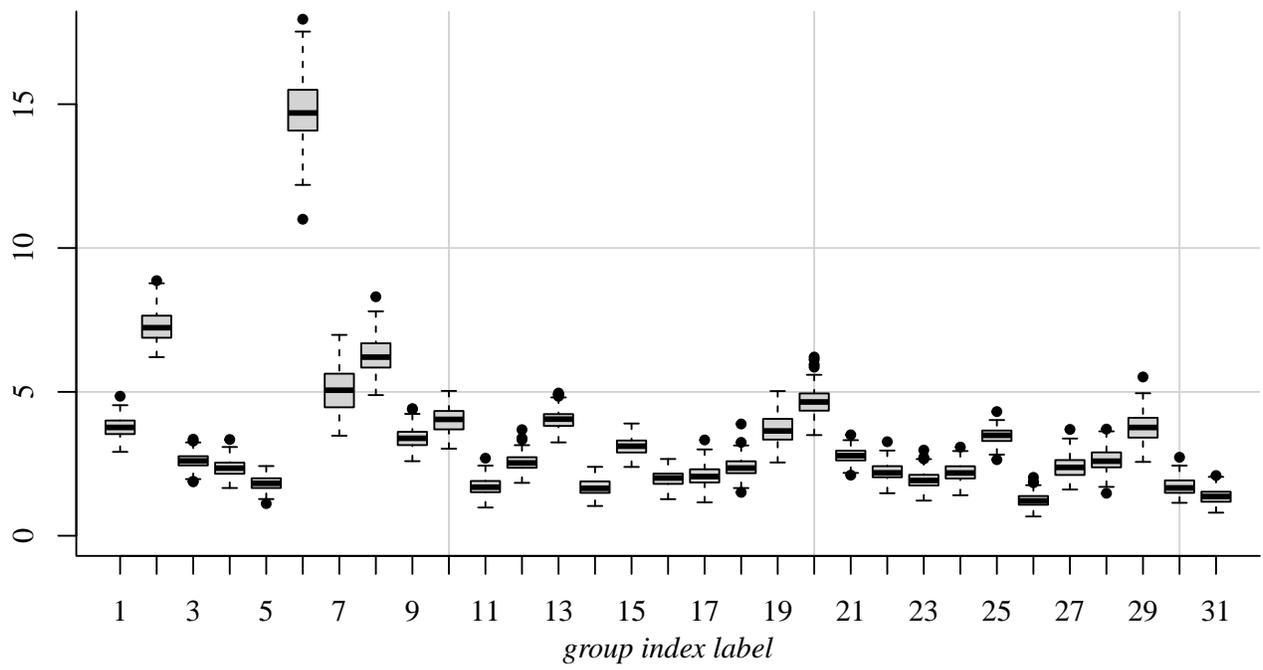

Figure 17: Boxplot for kernel parameters $\lambda_j$ associated to each group given the posterior modal partition; Model G.

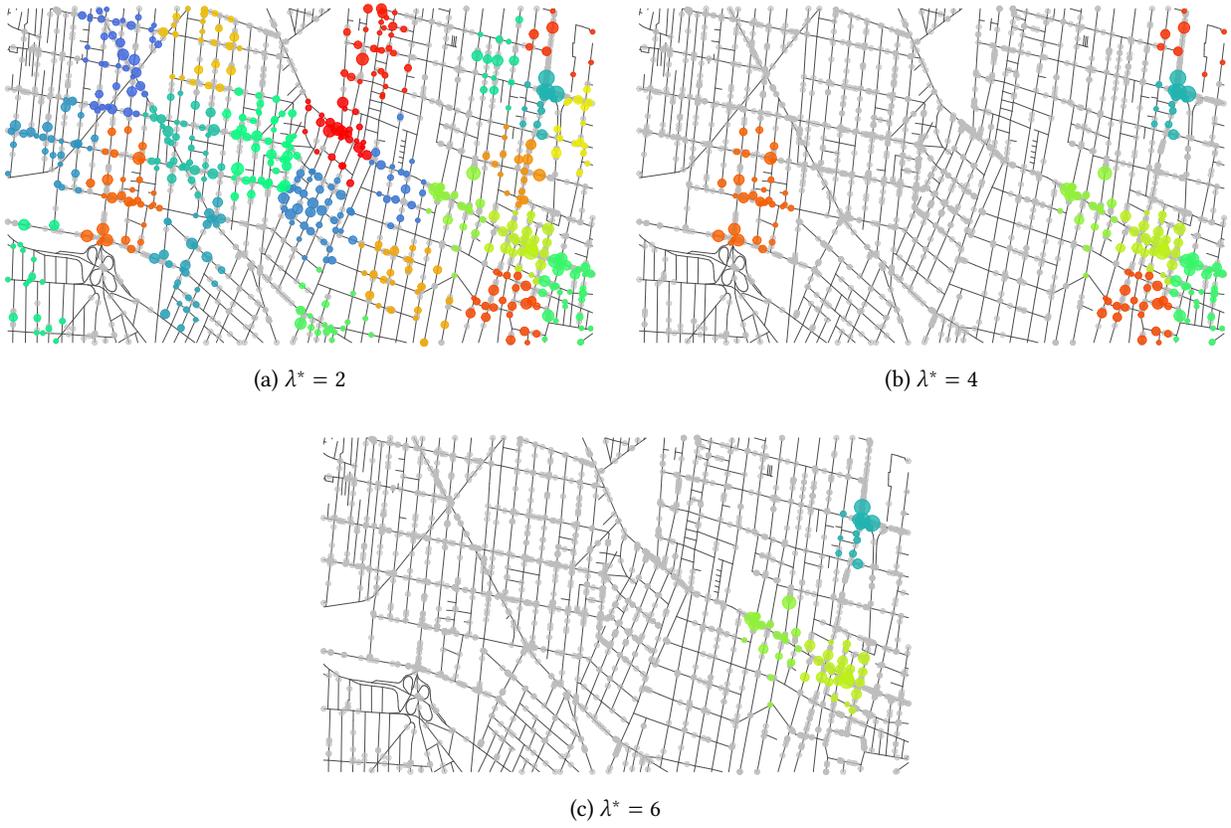

Figure 18: Restricted clustering, based on the modal partition, where posterior mean intensities, $\bar{\lambda}_j$, are above different values of $\lambda^*$: 2, 4 and 6; Model G.

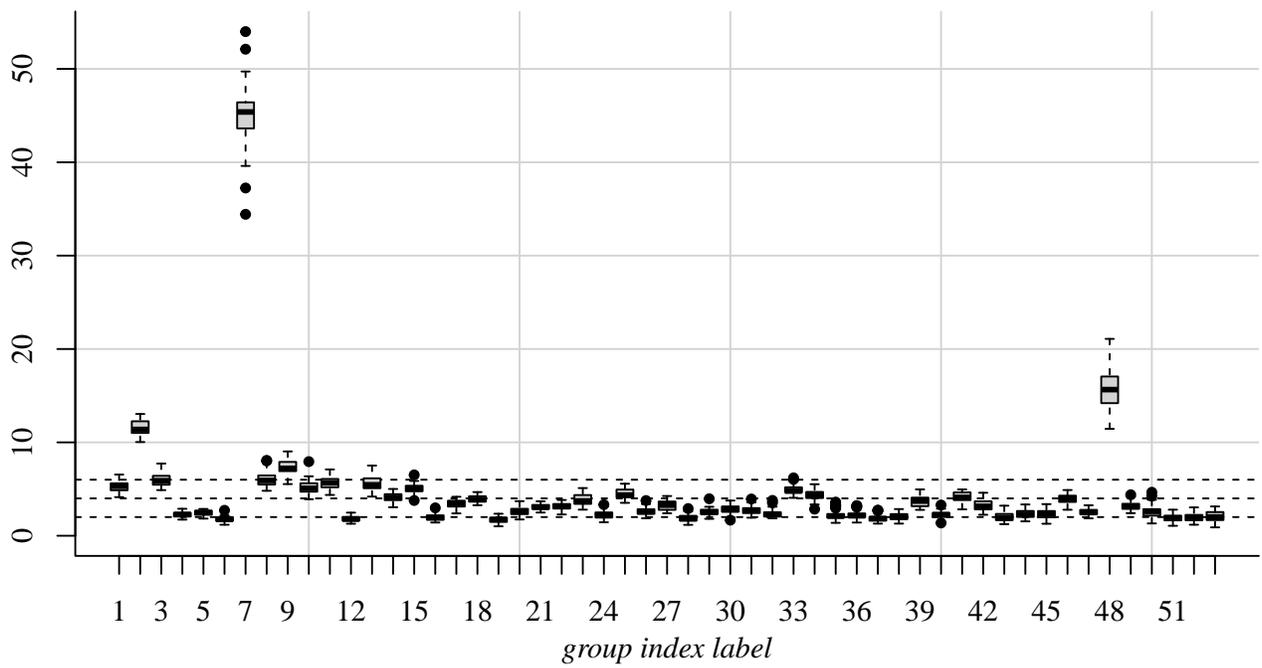

Figure 19: Boxplot for kernel parameters $\lambda_j$ associated to each group given the posterior modal partition; Model H.

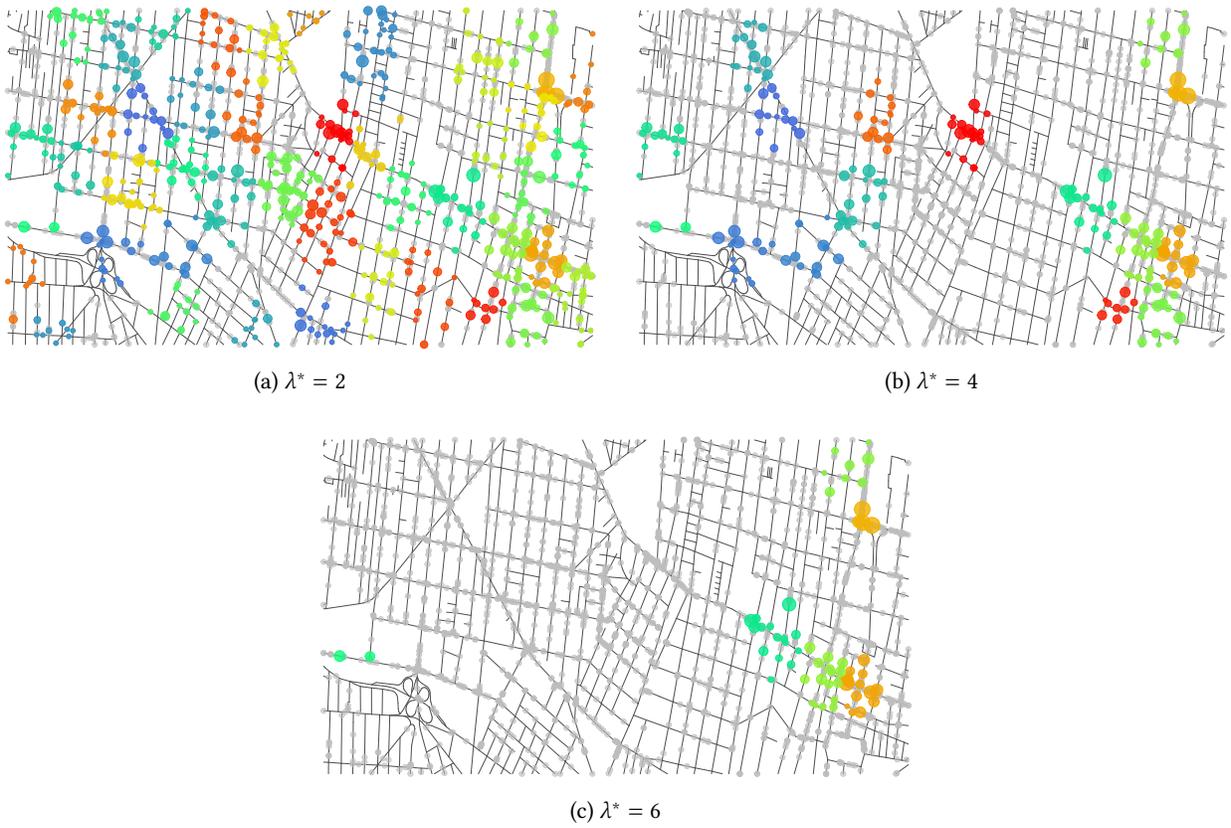

Figure 20: Restricted clustering, based on the modal partition, where posterior mean intensities, $\bar{\lambda}_j$, are above different values of $\lambda^*$: 2, 4 and 6; Model H.

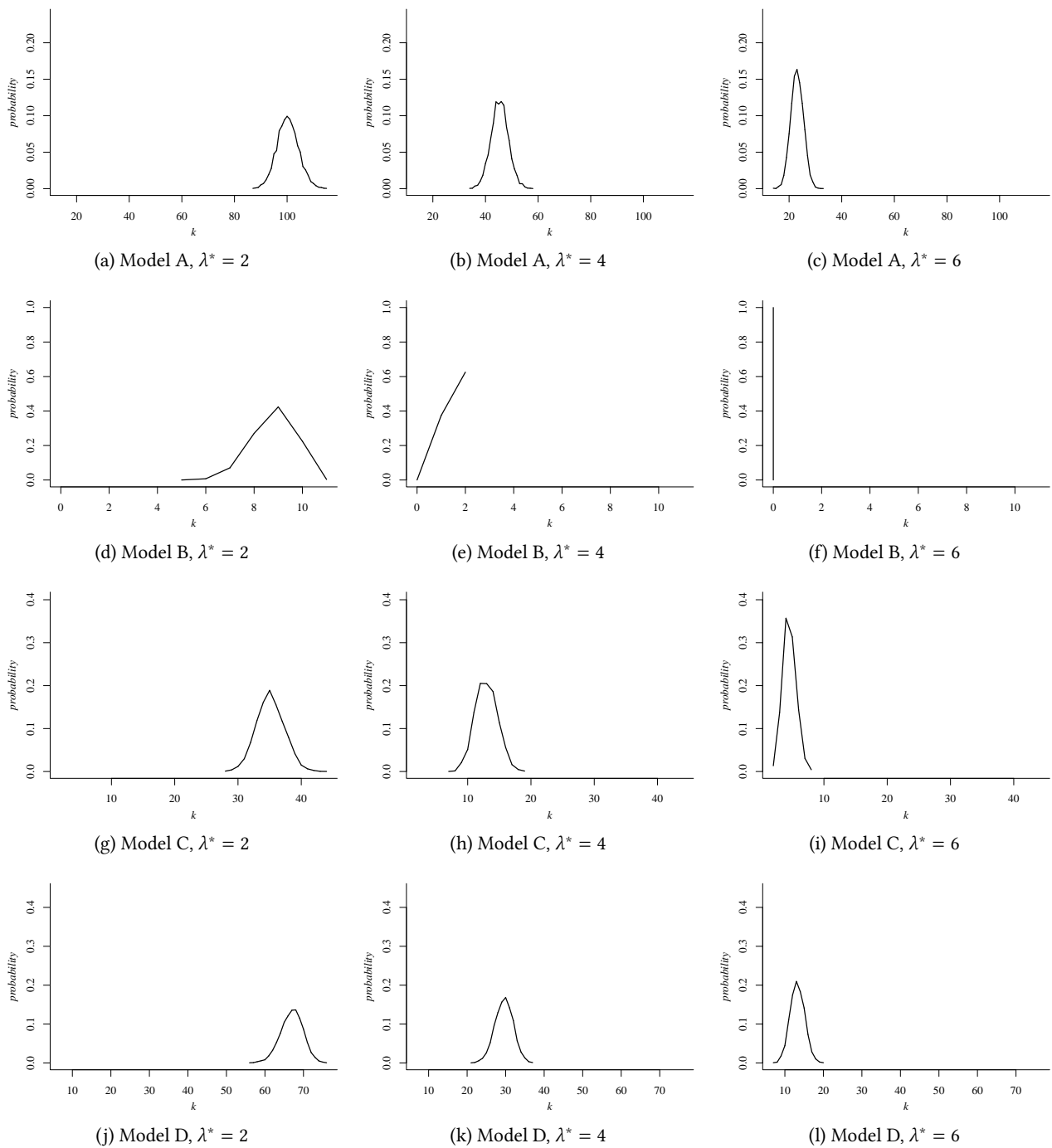

Figure 21: Conditional posterior distribution for the number of groups.

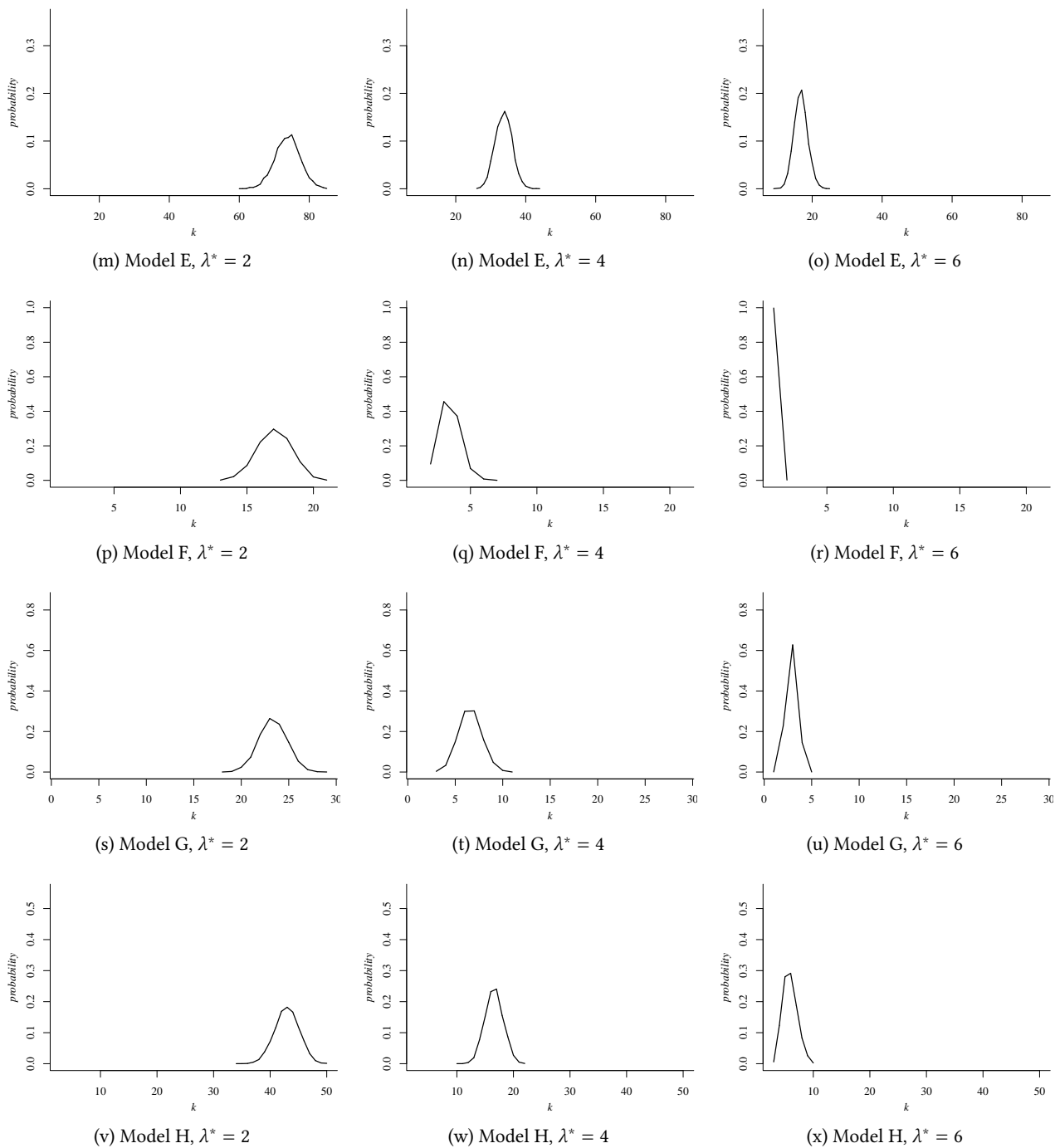

Figure 21: Conditional posterior distribution for the number of groups, continued.